\documentclass[letterpaper,twocolumn,10pt]{article}
\usepackage{usenix2019_v3}
\usepackage{filecontents}
\usepackage{tikz}

\usepackage{authblk}

\usepackage{lipsum}
\usepackage{fancyhdr}
\usepackage{epsfig,endnotes,color}
\usepackage{graphicx}
\usepackage{amsmath,bm}
\usepackage{graphicx}
\usepackage{CJKutf8}
\usepackage{subfigure}
\usepackage{rotating} 
\usepackage{lscape, latexsym, amssymb, multirow}
\usepackage{multirow}
\usepackage{mathtools, bbm, color}
\usepackage{booktabs}
\usepackage{rotating}
\usepackage{amsthm,mathrsfs,amsfonts,dsfont}
\usepackage{enumerate}
\usepackage{hhline}
\usepackage{enumitem}
\usepackage{tabu}
\usepackage{colortbl} 
\usepackage{enumerate}
\usepackage{caption}
\usepackage{footnote}
\usepackage{color}
\usepackage{colortbl}
\usepackage{pifont}
\usepackage{threeparttable}
\usepackage[framemethod=TikZ]{mdframed}
\usepackage{etoolbox,xspace}
\usepackage{algorithm}
\usepackage{algpseudocode}
\usepackage{bookmark}
\usepackage{longtable}
\usepackage{supertabular}

\usepackage{listings}

\usepackage{marvosym}
\usepackage{titlesec}

\hypersetup{colorlinks,linkcolor={black},citecolor={black},urlcolor={black}}  

\hypersetup{
	colorlinks=true,
	linkcolor=blue,
	filecolor=blue,      
	urlcolor=blue,
	citecolor=blue,
}

\definecolor{orange}{rgb}{1,0.5,0}

\newcommand{\ignore}[1]{}

\newcommand{\revised}[1]{}
\newcommand\comment[1]{}

\newcommand{\system}{{\sc UniFuzz}\xspace}

\newif\ifsubmit

\submitfalse
\ifsubmit
\newcommand{\ting}[1]{}
\else
\newcommand{\ting}[1]{\textcolor{red}{(ting: #1)}}
\fi

\begin{document}
\title{\Large \bf \system: A Holistic and Pragmatic Metrics-Driven Platform for Evaluating Fuzzers}

\author[$1$]{Yuwei Li}
\author[$1, 2$]{Shouling Ji}
\author[$1$]{Yuan Chen}
\author[$4$]{Sizhuang Liang}
\author[$5$]{Wei-Han Lee} 
\author[$1$]{Yueyao Chen}

\author[$1$]{Chenyang Lyu}
\author[$1, 3$]{Chunming Wu}
\author[$4$]{Raheem Beyah}
\author[$2, 1$]{Peng Cheng}
\author[$6$]{Kangjie Lu}
\author[$7$]{Ting Wang}

\affil[ ]{$^1${\small  Zhejiang University}, 
$^2${\small Zhejiang University NGICS Platform}, 
$^3${\small Zhejiang Lab, Hangzhou, China},

$^4${\small Georgia Institute of Technology},
$^5${\small IBM Research},
$^6${\small University of Minnesota},
$^7${\small Pennsylvania State University}
}

\affil[ ]{{\small E-mails: ~liyuwei@zju.edu.cn,
~~sji@zju.edu.cn, ~~chenyuan@zju.edu.cn, ~~liangsizhuang@gatech.edu, ~~wei-han.lee1@ibm.com, ~~coffee.ki.hy@gmail.com,
~~puppet@zju.edu.cn,
~~wuchunming@zju.edu.cn,
~~rbeyah@ece.gatech.edu,
~~saodiseng@zju.edu.cn,
~~kjlu@umn.edu,
~~inbox.ting@gmail.com.}}

\maketitle
\newcommand\blfootnote[1]{%
\begingroup
\renewcommand\thefootnote{}\footnote{#1}%
\addtocounter{footnote}{-1}%
\endgroup
}
\blfootnote{Yuwei Li and Shouling Ji are the co-first authors. 
Shouling Ji and Chunming Wu are the  co-corresponding authors.}
\footrule

\begin{abstract}
A flurry of fuzzing tools (fuzzers) have been proposed in the literature, 
aiming at detecting software vulnerabilities effectively and efficiently. 
To date, it is however still challenging to compare fuzzers due to the inconsistency of the benchmarks, performance metrics, and/or environments for evaluation, which buries the useful insights and thus impedes the discovery of promising fuzzing primitives.
In this paper, 
we design and develop \system, 
an open-source and metrics-driven platform for assessing fuzzers in a comprehensive and quantitative manner.
Specifically, \system to date has incorporated 35
usable fuzzers, 
a benchmark of 20 real-world programs, 
and six categories of performance metrics. 
We first systematically study the usability of existing fuzzers, 
find and fix a number of flaws,
and integrate them into \system.
Based on the study, we propose a collection of pragmatic performance
metrics to evaluate fuzzers from six complementary perspectives. 
Using \system,
we conduct in-depth evaluations of several prominent fuzzers including 
AFL~\cite{AFL}, 
AFLFast~\cite{Pham2016Coverage}, 
Angora~\cite{Chen2018Angora}, 
Honggfuzz~\cite{honggfuzz}, 
MOPT~\cite{mopt},
QSYM~\cite{yun2018qsym}, 
T-Fuzz~\cite{T-Fuzz} and VUzzer64~\cite{Rawat2017VUzzer}. 
We find that  none of them outperforms the others across all the target programs, and that using a single metric to assess the performance of a fuzzer may lead to unilateral conclusions,
which demonstrates the significance of comprehensive metrics.
Moreover, 
we identify and investigate previously overlooked factors that may significantly affect a fuzzer's performance, 
including \emph{instrumentation methods} and \emph{crash analysis tools}.
Our empirical results show that they are critical to the evaluation of a fuzzer.
We hope that our findings can shed light on reliable fuzzing evaluation, 
so that we can discover promising fuzzing primitives to effectively
facilitate fuzzer designs in the future. 

\end{abstract}

\section{Introduction}
Fuzzing is a software-testing technique that detects vulnerabilities by executing target programs with a large amount of abnormal or random test cases.
In recent years, 
a plethora of fuzzing related works have emerged in both industry and academia. 
In industry, 
major software vendors 
such as Google \cite{Google} and Microsoft \cite{MS}
leverage fuzzing techniques to help detect bugs in their products. 
On the other hand, GitHub \cite{github} hosts
more than 2,000 fuzzing related repositories.
In academia, 
over 200 fuzzing related research papers have been published since 2010, 
according to DBLP \cite{dblp}.

Despite the rapid development of fuzzing techniques,
there are several open questions that need to be addressed.
(1) \emph{How do these fuzzers perform in practice?}
(2) \emph{How  to compare different fuzzers under a fair and comprehensive set of performance metrics?}
(3) \emph{Which fuzzing primitives or techniques are 
promising and should be promoted?}
However,
previous works fail to answer these questions for the following reasons.
First,
many existing works do not conduct appropriate and sufficient experiments to provide trustworthy results.
For instance,
it is common to see that insufficient repetitions in the experiments make the results random and unreliable~\cite{klees2018evaluating}.
In addition,
many fuzzing works,
when comparing their methods with others,
directly use previously reported results without re-running the experiments~\cite{li2017steelix, T-Fuzz},
which is unfair as their experimental environments (e.g., CPU, memory) are different. 
Second,
the evaluations of existing fuzzers are often biased due to the lack of
uniform benchmarks.
The choices of target programs in different fuzzing papers vary widely.
Therefore, 
it is possible that the proposed fuzzers have preference over the selected programs.
Third,
the existing metrics are not suitable nor comprehensive for evaluating fuzzers.
It is inappropriate to only utilize the number of unique crashes to represent a fuzzer's capability of finding bugs,
as there is often a huge discrepancy between the number of unique crashes and the number of unique bugs~\cite{klees2018evaluating}.
In addition,
most existing fuzzing works do not evaluate the consumption of computing resources of the fuzzers.
Therefore,
there is an urgent need to conduct comprehensive and pragmatic evaluations for the state-of-the-art fuzzers on a uniform platform.

Conducting comprehensive and pragmatic evaluations of fuzzers entails overcoming multiple important challenges.
First,
although many fuzzers have been open sourced,
their usability in practice is often limited, as reported by recent
research~\cite{T-Fuzz, zhu2019feature}, 
which results in 
reproducibility issues, impeding comparison.
Thus,
it is necessary to test and enhance fuzzers' usability.
Second,
the evaluation of fuzzers should be conducted on pragmatic benchmark programs. 
Existing benchmark programs are not satisfactory \cite{klees2018evaluating}.
A reliable evaluation of fuzzers thus calls for pragmatic benchmarks.
Third,
the assessment must be conducted based on a comprehensive set of performance metrics.
Nevertheless,
existing metrics are insufficient and rough, leading to incomplete assessments.
Thus,
it is important to augment the performance metrics for comprehensive evaluation.

To address the above challenges,
we design and implement \system \cite{unifuzz},
an open-source, holistic and pragmatic metrics-driven platform for 
evaluating fuzzers.
In summary, we make the following main contributions.

\textbf{1) An Open-source and Pragmatic Metrics-driven Platform.} 
We design and implement \system,
the first open-source platform for evaluating fuzzers in a
comprehensive and quantitative manner,
which to date has incorporated 35 popular fuzzers, a benchmark suite of 20 real-world programs,
and six categories of performance metrics.
For each of the 35 fuzzers,
we test its usability and provide a Dockerfile for easy installation and deployment.
In addition,
we find and fix (partially) more than 15 flaws,
which have been reported to their developers.
For the 20 real-world benchmark programs,
\system provides all necessary side information such as software installation and command arguments to ensure their usability. 
Furthermore, 
we implement tools in \system to facilitate the crash analysis process including triaging crashes into bugs, matching with the corresponding CVEs, 
and analyzing the severity of the bugs, etc.
Specifically,
we develop a \emph{CVE keywords database} that includes the CVEs for the \system benchmark programs,
which can significantly reduce the human efforts in CVE matching.
We also propose a collection of performance metrics in six categories:
\emph{quantity of unique bugs},
\emph{quality  of bugs},
\emph{speed of finding bugs},
\emph{stability of finding bugs},
\emph{coverage} and \emph{overhead},
which can be used to assess a fuzzer's performance comprehensively.

\textbf{2) Extensive Evaluations of Fuzzers.} 
Leveraging \system, we conduct extensive experiments to compare eight prominent coverage-based fuzzers.
The experimental results show that no fuzzer outperforms the others on all the tested benchmark programs, which are very different from the conclusions in their papers.
This observation reveals that subjectivity and bias may exist in the evaluations of previous fuzzing works.
Moreover, the experimental results reflect that using a single metric to evaluate fuzzers may lead to unilateral conclusions, which demonstrates the importance of using comprehensive metrics to evaluate the fuzzers.

\textbf{3) New Findings and Insights for Future Fuzzing.}
From the evaluations, we gain important insights and findings for future fuzzing research. 
For example, we find previously unaccounted factors that can significantly affect the performance of fuzzers, e.g., 
\emph{instrumentation methods} and \emph{crash analysis tools}.
The results demonstrate that even small changes of these factors can have a significant impact on the assessment of fuzzers.
Therefore, fuzzing experiments should be conducted in a more rigorous and precise way to provide more reliable results.

\section{Motivation of \system}\label{sub-motivation}
To assess the performance of existing fuzzers and to enlighten the design of new ones, 
it is crucial to conduct in-depth comparative studies of different fuzzers.
Unfortunately, 
there are many challenges for conducting such comprehensive evaluations on fuzzers as follows, which motivate the design of \system.

\textbf{Usability Issues of Existing Fuzzers.}
Whether the existing implementation of fuzzers works in practice is often questionable.
First, some fuzzers may be difficult or complicated to be used directly.
For instance, Zhu et al. \cite{zhu2019feature} stated that they could not appropriately run Driller \cite{stephens2016driller}, T-Fuzz \cite {T-Fuzz} and VUzzer \cite{Rawat2017VUzzer}.
Second, we find that there are numerous flaws (e.g., incorrect judgment on crash, abnormal behaviors during the fuzzing process) with the implementation of many fuzzers,
which may cause negative impacts on their performance.
Therefore,
it is necessary to test the usability of existing fuzzers and call for more community efforts to enhance fuzzers' usability in practice.
We provide the detailed analysis of the flaws of several popular fuzzers on the \system open-source platform \cite{unifuzz} due to space limitation.

\textbf{Lack of Pragmatic Real-World Benchmark Programs.}
Benchmark programs are fundamental for evaluating the performance of fuzzers,
which should be carefully designed such that a fuzzer can be evaluated in a fair manner.
Thus,  good benchmark programs should have the following characteristics.
First,  they should have similar features as the real-world programs,
and these features include coding styles, sizes and vulnerabilities.
In this way,   a fuzzer's performance on these benchmark programs can be more indicative.
Second,  to provide comprehensive evaluations, benchmark programs should exhibit a diversity of functionalities,  sizes, vulnerability types, etc.
Third, from the perspective of conducting practical assessments on fuzzers' capabilities in discovering bugs, each benchmark program should contain at least one vulnerability that can be found within a reasonable amount of time, which implies two important properties of a pragmatic benchmark. 
(1) The program should contain at least one bug. Otherwise, it cannot effectively distinguish the capabilities of fuzzers in discovering bugs.
(2) The difficulty in discovering a bug should be reasonable\footnote{\scriptsize Note that the difficulty is relative to the state-of-the-art fuzzers. With the development of the fuzzers, the new benchmarks with higher difficulty need to be proposed.}. 
Otherwise, it may cause unaffordable evaluation overhead.
For instance, a one-month fuzzing experiment for a single fuzzer on a single program with 30 repetitions requires 21,600 CPU hours, let alone  conducting a reliable and comprehensive evaluation with multiple benchmark programs and seed sets~\cite{klees2018evaluating}.
Fourth, 
the benchmark programs should be easy to use.
To this end, the developers should provide rich information of a benchmark program such as installation methods, command arguments, input types.
Moreover,
it would be better if the developers of the benchmarks can provide methods/tools for automatically analyzing the corresponding crash samples of benchmark programs.

Existing fuzzing benchmark programs can be grouped into two categories: synthetic programs and real-world programs.
Typical examples of synthetic benchmarks include LAVA-M \cite{Dolangavitt2016LAVA} and DARPA CGC \cite{DARPACGC}.
Typical examples of real-world programs are \texttt{exiv2}, \texttt{mp3gain}, etc., which are Linux open-source programs with several vulnerabilities.
However, existing benchmark programs, both synthetic and real-world are not satisfactory \cite{klees2018evaluating}.

The existing synthetic benchmarks usually are small in size,  and the artificial bugs are designed and injected following some relatively simple mechanisms.
Thus, the developer of a fuzzer may improve its performance by understanding the bug-injecting patterns and the mechanisms, and the evaluation results can be biased.
As a result, 
fuzzers that have good performance on these synthetic benchmark programs may not work well on the real-world programs.

The existing real-world benchmark programs are not satisfactory as
well due to the following issues.
First,
we still lack standard and sufficient real-world benchmark 
programs, and existing fuzzers are usually evaluated on self-chosen programs, 
which may cause evaluation bias. 
Second, 
the real-world programs are not as convenient as the synthetic programs on validating bugs due to the lack of clear indicators of bug triggering.
For example,
existing works usually triage crashes and filter vulnerabilities by leveraging different tools such as AddressSanitizer (ASan) \cite{AS} and GDB \cite{gdb}. 
However,
due to their own limitations and inconsistency between different tools,
these different crash triage methods may cause bias as well.
Moreover,
many papers either state that they validate the corresponding CVEs manually~\cite{ognawala2019compositional, Pham2016Coverage, chen2014matryoshka} or do not mention how they validate the CVEs. %
Nevertheless,
the manual validation process is time-consuming and tedious,
which may also cause bias and mistakes.
All the issues call for the development of a suite of diverse and pragmatic benchmarks as well as automatic tools to analyze crashes.

\textbf{Lack of Proper and Comprehensive Performance Metrics.}
Most previous works usually evaluate fuzzers using the three de facto metrics:
the number of unique crashes,
the number of unique bugs,
and the coverage.
However,
these metrics alone often fail to fully account for a fuzzer's performance.
For instance,
solely relying on the number of unique crashes 
\cite{CollAFL, Pham2016Coverage, Rawat2017VUzzer}
may lead to misleading conclusions,
as more unique crashes do not definitely represent more unique bugs \cite{klees2018evaluating}.
Further,
in addition to the number of unique bugs, 
the quality of bugs is also an important metric that needs to be taken into consideration.
For example,
when two fuzzers find a similar number of bugs in the same time,
it is inappropriate to draw the conclusion that the two fuzzers have similar performance, 
if the bugs found by one fuzzer are rarer or more dangerous.
Finally,
overhead is also an important metric.
The number of bugs found by fuzzer \emph{A} may be twice as many as those found by fuzzer \emph{B}, 
but it might be improper to consider fuzzer \emph{A} performs better when it costs hundreds of times of computing resources than fuzzer \emph{B}.
Therefore, we need to enhance the metrics, so that they complement each other and provide comprehensive and reliable evaluations for fuzzers.

\section{Design of \system}\label{overview}
\begin{figure}[tb]
\centering
\includegraphics[width=3.5in,bb=0 0 310 190]{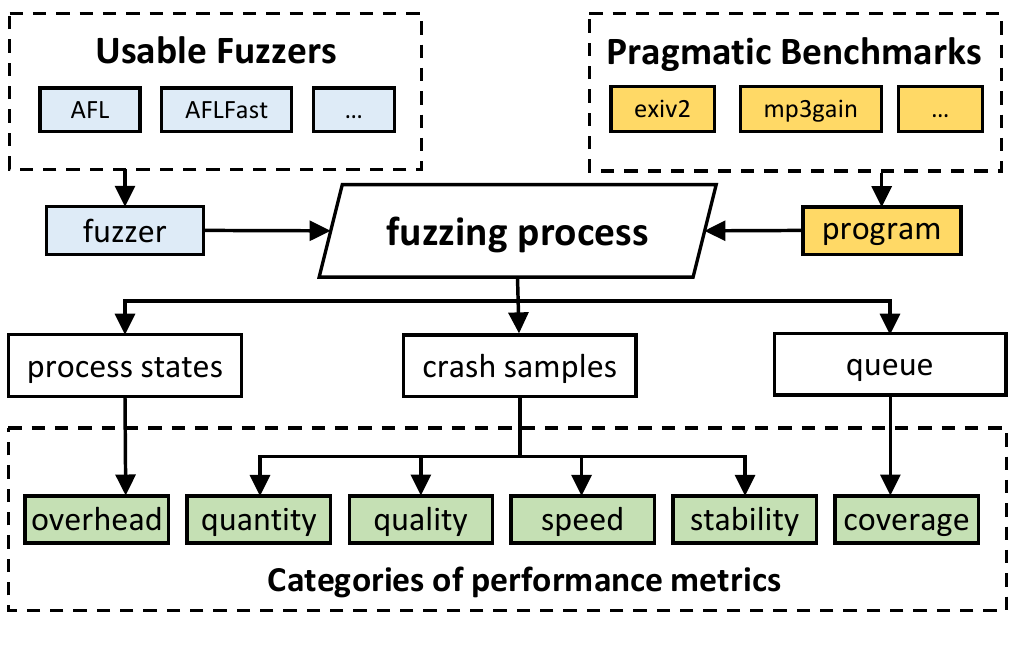}
\caption{Overview of \system.}
\label{fig-sys-overview}
\end{figure}

To address the challenges discussed in Section \ref{sub-motivation},
we design and implement \system,
an open-source platform for evaluating fuzzers.
Figure \ref{fig-sys-overview} presents an overview of \system,
which mainly consists of three components:
\emph{usable fuzzers},
\emph{pragmatic benchmarks}, and \emph{performance metrics}.

\subsection{Usable Fuzzers}
\system to date has incorporated 35 usable  fuzzers including
AFL~\cite{AFL},
AFLFast~\cite{Pham2016Coverage},
AFLGo~\cite{bohme2017directed},
AFLPIN~\cite{AFLPIN},
AFLSmart~\cite{AFLSmart},
Angora~\cite{Chen2018Angora},
CodeAlchemist~\cite{han2019codealchemist},
Driller~\cite{stephens2016driller},
Domato~\cite{domato},
Dharma~\cite{dharma},
Eclipser~\cite{eclipser},
FairFuzz~\cite{lemieux2018fairfuzz},
Fuzzilli~\cite{Fuzzilli},
Grammarinator~\cite{hodovan2018grammarinator},
Honggfuzz~\cite{honggfuzz},
Jsfuzz~\cite{jsfuzz},
jsfunfuzz~\cite{jsfunfuzz},
LearnAFL~\cite{learnafl},
MoonLight~\cite{hayes2019moonlight},
MOPT~\cite{mopt},
NAUTILUS~\cite{nautilus},
NEUZZ~\cite{neuzz},
NEZHA~\cite{nezha},
Orthrus~\cite{shastry2017static},
Peach~\cite{Peach},
PTfuzz~\cite{zhang2018ptfuzz},
QSYM~\cite{yun2018qsym},
QuickFuzz~\cite{grieco2016quickfuzz},
radamsa~\cite{radamsa},
slowfuzz~\cite{petsios2017slowfuzz},
Superion~\cite{superion},
T-Fuzz~\cite{T-Fuzz},
VUzzer~\cite{Rawat2017VUzzer},
VUzzer64~\cite{Rawat2017VUzzer} and
zzuf~\cite{zzuf}.
The types of incorporated fuzzers are diverse, including grammar-based, mutation-based, directed and coverage-based fuzzers.
Table~\ref{tab-fuzzer} presents the detailed information of the usable fuzzers incorporated in \system.
In order to test the usability of these fuzzers,
we manually build and test each of these fuzzers.
During this process,
we find many design and implementation flaws in these fuzzers.
Up to date,
we have found more than 15 serious flaws among these fuzzers and reported them to the developers.
With our help, some of these flaws have been promptly fixed and released.
A more detailed description of these issues is presented in the \system open-source platform \cite{unifuzz}. %
For each fuzzer in Table~\ref{tab-fuzzer},
we also implement a Dockerfile for installing and using it conveniently in a Docker container.
We choose to conduct fuzzing experiments in a Docker container for the following reasons.
First, compared with conducting fuzzing on a physical machine, Docker is more convenient for resource allocation and isolation, which can provide fair fuzzing evaluations.
Second, compared with virtual machines, Docker is lighter-weight and costs less computing resources.
Thus, with limited resources, users are able to conduct more fuzzing experiments simultaneously when using Docker.
In addition, Docker can be operated and managed more conveniently.
In addition to testing the usability of these fuzzers and making them available, we conduct comprehensive evaluations on eight prominent  coverage-based fuzzers, with the details presented in Section~\ref{general}.

\begin{table}[!t]
	\footnotesize
	\centering
	\caption{The fuzzers incorporated in \system.}\label{tab-fuzzer}
	\begin{threeparttable}
		\setlength{\tabcolsep}{2pt}
		\scalebox{0.9}{
			\begin{tabular}{lccccc}
				\hline
				\textbf{Fuzzer } & \textbf{Mutation/Generation} & \textbf{Directed/Coverage} & \textbf{Target} \\
				\hline
				AFL~\cite{AFL}  & M   & C & S/B \tnote{1}  \\ 
				AFLFast~\cite{Pham2016Coverage} & M  & C & S/B \\ 
				AFLGo~\cite{bohme2017directed} & M  & D & S     \\ 
				AFLPIN~\cite{AFLPIN}    & M     & C & B   \\ 
				AFLSmart~\cite{AFLSmart} & M &C & S/B \\
				Angora~\cite{Chen2018Angora} & M  & C & S/B                 \\ 
				CodeAlchemist~\cite{han2019codealchemist} & G & n.a. & B \\
				Driller~\cite{stephens2016driller}  & M & C & B     \\
				Domato~\cite{domato}   & G   & n.a. & B    \\  
				Dharma~\cite{dharma}    & G  & n.a. & B      \\
				Eclipser~\cite{eclipser} & M & C & S \\
				FairFuzz~\cite{lemieux2018fairfuzz}  & M  & C & S    \\
				Fuzzilli~\cite{Fuzzilli} & M & C & S \\
				Grammarinator~\cite{hodovan2018grammarinator}  & G      & n.a. & B          \\ 
				Honggfuzz~\cite{honggfuzz}   & M   & C & S   \\
				Jsfuzz~\cite{jsfuzz} & M & C & S \\
				jsfunfuzz~\cite{jsfunfuzz} & G & n.a. & B \\
				LearnAFL~\cite{learnafl} & M &C  & S \\
				MoonLight~\cite{hayes2019moonlight} & n.a. & n.a. & n.a. \\
				MOPT~\cite{mopt} & M & C & S/B \\
				NAUTILUS~\cite{nautilus}  &  G+M & C & S \\ 
				NEUZZ~\cite{neuzz} & M & C & S \\
				NEZHA~\cite{nezha} & M  & C   & L \tnote{2}  \\ 
				Orthrus~\cite{shastry2017static}  & n.a.   & n.a. & n.a.  \\
				Peach~\cite{Peach}       & G    & n.a.   & B     \\ 
				PTfuzz~\cite{zhang2018ptfuzz}   & M   & C   & S   \\ 
				QSYM~\cite{yun2018qsym}   & M  & C  & B  \\
				QuickFuzz~\cite{grieco2016quickfuzz}  & G+M   & n.a.  & B    \\ 
				radamsa~\cite{radamsa}  & M  & C  & B  \\ 
				slowfuzz~\cite{petsios2017slowfuzz} & M  & n.a. & L \\ 
				Superion~\cite{superion} & G+M & C & S \\
				T-Fuzz~\cite{T-Fuzz}  & M   & C  & S  \\ 
				VUzzer~\cite{Rawat2017VUzzer}  & M   & C   & B  \\ 
				VUzzer64~\cite{Rawat2017VUzzer}   & M   & C    & B  \\ 
				zzuf~\cite{zzuf}                      & M   & n.a.   & B     \\
				\hline 
		\end{tabular}}
		\begin{tablenotes}
			\footnotesize
			\item[1] S: source code, B: binary. 
			\item[2] L: user needs to write libFuzzer code.
		\end{tablenotes}
	\end{threeparttable}
\end{table}

\subsection{Pragmatic Benchmarks}\label{benchmark}\label{sub-prog}
According to Section~\ref{sub-motivation},
pragmatic benchmark programs should have the following properties:
(1) \emph{similar to the real-world programs}, including coding styles, sizes, and vulnerabilities, etc.
(2) \emph{comprehensive}, which are various in terms of functionalities, sizes, and vulnerability types, etc.
(3) \emph{practical}, which means at least one bug should be found in a reasonable amount of time.
(4) \emph{conveniently to be used}, which means the users can easily use the benchmark programs and get the evaluation results.
Following the above principles,
we construct a pragmatic benchmark suite that consists of 20 real-world programs for evaluating fuzzers as shown in Table \ref{tab-bench}.
Specifically,
\system provides detailed and comprehensive information of each program including the version, 
size,
installation information,
input type and command arguments, etc., 
to ensure the usability.
For each program,
\system provides its source code and a Dockerfile for installing and using it.
Furthermore,
\system provides effective tools to analyze the corresponding crash samples of a target program conveniently.
The analyses include but are not limited to:
(1) de-duplicating and triaging the crash samples into bugs;
(2) matching the crash samples into the corresponding CVEs and
(3) analyzing the severity of the bugs triggered by the crash samples.
It is worth noting that we do not modify the benchmark programs.
As a result, the raw features of the real-world programs are preserved.
Instead,
we focus on how to select these programs and developing tools for analyzing the experimental results conveniently.
Next,
we describe the details of program selection and  the crash analysis methods.

\begin{table}[tb]
\footnotesize
\centering
\renewcommand\arraystretch{1.2}
\caption{The real-world programs of the \system benchmark. @@  represents the input file.}
\label{tab-bench}
\setlength{\tabcolsep}{2mm}{
\scalebox{0.75}{
\begin{tabular}{l|llllll}
\hline
\textbf{Type} & \textbf{Program} & \textbf{Version}   & \textbf{Arguments}  \\
\hline 
\multirow{5}{*}{Image} & \texttt{exiv2} &0.26         & @@       \\
\cline{2-4}
& \texttt{gdk-pixbuf-pixdata (gdk)}   
& gdk-pixbuf 2.31.1       &  @@ /dev/null      \\
\cline{2-4}
& \texttt{imginfo}           & jasper 2.0.12            & -f @@       \\
\cline{2-4}
& \texttt{jhead}             & 3.00                     & @@       \\
\cline{2-4}
& \texttt{tiffsplit}         & libtiff 3.9.7            & @@       \\
\cline{1-4}
\multirow{3}{*}{Audio}  & \texttt{lame} & lame 3.99.5              &   @@ /dev/null \\
\cline{2-4}
& \texttt{mp3gain}           & 1.5.2-r2                 & @@  \\
\cline{2-4}
& \texttt{wav2swf}           & swftools 0.9.2            & -o /dev/null @@ \\
\cline{1-4}
\multirow{5}{*}{Video}   & \multirow{3}{*}{\texttt{ffmpeg}}        & \multirow{3}{*}{4.0.1}  & (-y -i @@ -c:v $\backslash$ \\
&   &    & mpeg4 -c:a copy -f $\backslash$ \\
&   &    & mp4 /dev/null)\\
\cline{2-4}
& \texttt{flvmeta}             & 1.2.1                  & @@ \\
\cline{2-4}
& \texttt{mp42aac}             & Bento4 1.5.1-628         & @@ /dev/null  \\
\cline{1-4}
\multirow{6}{*}{Text}   & \texttt{cflow} & 1.6                     & @@\\
\cline{2-4}
& \texttt{infotocap}          & ncurses 6.1             &  -o /dev/null @@ \\
\cline{2-4}
& \texttt{jq}                 & 1.5                     & . @@ \\
\cline{2-4}
& \texttt{mujs}               & 1.0.2                   &  @@ \\
\cline{2-4}
& \texttt{pdftotext}          & xpdf 4.00               & @@ /dev/null \\
\cline{2-4}
& \texttt{sqlite3}            & 3.8.9                   &  (stdin) \\
\cline{1-4}
\multirow{6}{*}{Binary}   & \multirow{5}{*}{\texttt{nm}}     & \multirow{5}{*}{binutils 5279478}  & (-A -a -l -S -s $\backslash$ \\				
&&& -\,-special-syms $\backslash$ \\ 
&&& -\,-synthetic $\backslash$  \\
&&& -\,-with-symbol-versions  $\backslash$ \\
&&& -D @@)  \\
\cline{2-4}
& \texttt{objdump}            & binutils 2.28           & -S @@ \\
\cline{1-4}
Network & \texttt{tcpdump} & 4.8.1 + libpcap 1.8.1   &  -e -vv -nr  @@ \\
\hline
\end{tabular}}}
\end{table}

\textbf{Programs Selection.}
In order to select suitable programs, we investigate fuzzing-related papers published on top conferences in information security and software engineering fields to find the real-world programs and the corresponding versions used in their evaluations\footnote{\scriptsize If a program is selected with multiple versions, we prefer to choose the one which has more vulnerabilities.}.
Based on the above process,
we finally select 20 real-world programs as shown in Table \ref{tab-bench}.
The selected programs cover six functionality types including: image, audio, video, text, binary and network packet processing
software.
In addition,
they cover various types of vulnerabilities including: 
\emph{heap buffer overflow},
\emph{stack overflow},
\emph{segmentation fault},
\emph{excessive memory allocation},
\emph{global buffer overflow},
\emph{stack buffer overflow},
\emph{memory leak},
\emph{free error},
\emph{float point exception},
\emph{alloc-dealloc mismatch},
\emph{memcpy parameter overlap},
\emph{use-after-free}, etc.
Therefore, these programs are able to provide a comprehensive evaluation on the performance of fuzzers.

\textbf{Triaging Crashes into Unique Bugs.}
Generally,
there are two main approaches for triaging crashes into unique bugs: 
one is based on analyzing the root cause of the bugs, 
and the other is based on analyzing the output results.
One common implementation of the first approach is to patch the program for each vulnerability based on the analysis of the root cause of the vulnerability \cite{klees2018evaluating}.
If crash file \emph{a} and crash file \emph{b} both trigger the bug of the target program, but neither does that on the bug-fixed one, they will be regarded as the same bug.
Although this approach seems to be able to provide accurate ground truth information of the benchmark programs,
the root cause analysis is hard~\cite{lahiri2015automatic, nguyen2013semfix} and there are many challenges in implementing this approach in practice.
(1) To provide all-side ground truth information of bugs in the program, it is required to access all the patched versions of the target program.
(2) Each patched version should only fix one unique bug without overlap.
Otherwise, it may cause huge false positives/negatives.  

The second approach is usually implemented by analyzing the output information when bugs are triggered.
Compared to the first approach, the second approach is more practical in implementation which can provide relatively fair evaluation results.
For instance, one commonly used method is leveraging tools such as ASan \cite{AS} to produce the stack trace information when a bug of the program is triggered, then the stack hash method \cite{molnar2009dynamic} can be used to extract \emph{N} stack frames to de-duplicate the bugs.
This approach may also cause false positives/negatives when choosing different values of \emph{N}.
Nevertheless, 
how to select the value of \emph{N} to provide results with the lowest false positives/negatives is a difficult research problem which has not been completely solved and is out-of-scope of this paper.
As a trade-off and guided by the previous work~\cite{klees2018evaluating, molnar2009dynamic},
we select \emph{N} as 3.
In addition, as different tools use various methods to detect bugs,
relying on a single tool may neglect certain types of bugs.
Therefore, to have a more precise detection result, we prioritize the output report produced by ASan \cite{AS} and use the output reports produced by other tools such as GDB \cite{gdb} as a supplement\footnote{\scriptsize We only use GDB to detect the bugs which cannot be found by ASan. 
Therefore, a crash sample can only be mapped with one unique bug at most.}.

\textbf{Matching CVEs.}
Common Vulnerabilities and Exposures (CVE) \cite{CVE} provides information of existing vulnerabilities.
Most existing fuzzing works evaluate their fuzzers' capability in finding bugs by leveraging CVE information \cite{CollAFL, Pham2016Coverage, T-Fuzz}.
Thus, it is important to figure out what and how many known/new CVEs (CVE vulnerabilities) are discovered by a fuzzer.
However, matching crash samples with the corresponding CVEs is time-consuming and tedious for the following reasons.
First, the description of each CVE is written in natural language without a well-defined structure.
Thus, it is difficult to extract the key information (e.g., vulnerable function names) from the description directly.
Second, although the references of each CVE may provide additional information such as the PoC (Proof-of-the-Concept) file that triggers a CVE and the output report generated by crash analysis tools, such information is usually incomplete or missing \cite{mu2018understanding}, which makes CVE matching even harder.
Moreover, the references are also unstructured.
Thus, human efforts are needed to figure out what content a reference link represents (e.g., the download link of a PoC file or the bug reports).
Third, as different tools may be leveraged to obtain the output report, it is difficult to match with different output reports directly.

In order to reduce the human efforts in matching CVEs, we construct a \emph{CVE keywords database} that includes the key information of the CVEs related to the \system benchmark programs.
This database can be leveraged to match the crashes with the corresponding CVEs conveniently.
Compared with the information provided on the official CVE website \cite{CVE}, the information in \emph{CVE keywords database} is better structured.
In \emph{CVE keywords database}, each benchmark program has a CVE table.
Each entry of the table represents the information of a CVE.
The primary key of each entry is the CVE ID, and the other attributes are the pivotal information of this CVE, including vulnerability type, vulnerable functions, vulnerable files, stack trace, the tool that generates the stack trace, etc.
Leveraging the \emph{CVE keywords database}, we implement a method that can conveniently generate the initial matching results.
Based on the \emph{CVE keywords database}, the CVE matching process is as follows.
(1) Compile the program with the corresponding tool (e.g., ASan) and execute the binary with the crash to obtain the output report.
(2) Extract the necessary information from the output report, which includes the stack trace, vulnerability type, vulnerable functions, vulnerable file names, etc.
(3) Match the extracted information with the CVE table of the program in \emph{CVE keywords database} and report the initial matching CVEs sorted by the number of matched keywords.
(4) Check the initial matching results manually to obtain the final matching results.
Note that  the official CVE website~\cite{CVE} has flaws and mistakes such as incomplete information~\cite{mu2018understanding} and overlapped CVEs \cite{overlap}.
On the other hand, it is possible that a 0-day vulnerability is found.
Thus, in this case,  it is necessary to conduct the last step to make the matching result more precise and accurate.

\emph{Discussion on the Ground Truth.}
In general, it is hard to obtain the complete  ground truth bugs for both the synthetic and the real-world programs due to the nature of bugs.
For the synthetic benchmarks, whether the other parts (except for the injected bugs) have bugs is unknown, which makes it hard to obtain the complete ground truth.
Similar for a real-world program, except for the already known bugs, whether it has new bugs is unknown.
Even though, we try our best to provide the information as accurate as possible for the benchmark in the following manners.
First, to mitigate the incompleteness issue, we collect as many crash samples as possible to detect the possible bugs in the benchmark programs.
Second, we use multiple tools to detect the bugs.
In addition to the eight coverage-based fuzzers, we combine three static analysis tools (Flawfinder \cite{Flawfinder}, RATS \cite{rats}, Clang Static Analyzer \cite{clang-static-analyzer}) with the directed fuzzer, AFLGo \cite{bohme2017directed}, to find more bugs of the \system benchmark.
The details of the detection results are presented in the \system open-source platform \cite{unifuzz},  due to space limitation.
Third, we analyze the bugs with multiple tools (i.e., ASan and GDB) to reduce the impact caused by the limitations of a single tool.

\subsection{Performance Metrics}\label{metrics}

To address the problem of lacking comprehensive and pragmatic performance metrics,
we systematically study the performance metrics of the existing fuzzing papers,
summarize and propose a set of metrics, 
which can be classified into six categories:
\emph{quantity of unique bugs},
\emph{quality of bugs},
\emph{speed of finding bugs},
\emph{stability of finding bugs},
\emph{coverage} and \emph{overhead}.
Each category represents a property of a fuzzer‘s performance,
and each property can be evaluated by many concrete metrics which are expandable.
For example,
when evaluating \emph{quantity of unique bugs},
we can leverage many concrete mathematical metrics such as $p$ value,  $\hat{A}_{12}$ score \cite{vargha2000critique}.
In the following,
we present concrete metrics for each category as suggestions to use in practical evaluations.

\textbf{Quantity of Unique Bugs.}
As there exists randomness with a fuzzing process,
a robust fuzzing experiment has to be repeated for multiple times to provide a more reliable result.
Therefore,
the quantitative metrics of unique bugs are based on statistical methods.
We focus on two important questions:
(1) \emph{how many times should a fuzzing experiment be repeated?}
and 
(2) \emph{what statistical metrics can provide reliable results?}
There are different opinions about these questions.
For question (1),
Klees et al. \cite {klees2018evaluating} suggested conducting each fuzz testing for 30 repetitions.
For question (2),
Klees et al. \cite{klees2018evaluating} stated that general statistical metrics such as \emph{mean},
\emph{median} and \emph {variance} may result in misleading conclusions.
Besides, 
they highly recommended to use statistical tests to calculate the $p$ value to determine whether there is a statistically significant difference between the two fuzzers' performance.
Specifically,
they suggested using the Mann-Whitney U test as the statistical test method instead of other methods.
The reason is that the Mann-Whitney U test is non-parametric which makes no assumption on the distribution of the population (as the distribution of fuzzing results, e.g., the number of unique bugs in all repetitions is still unknown),
whereas some other methods are stricter.
For instance, 
$t$-test assumes that the two populations must obey normal distributions and have the same variance.
However, 
there are some different viewpoints about statistical tests.
For example, Nuzzo \cite{nuzzo2014scientific} pointed out that the $p$ value is not as reliable as many scientists assume, and Wasserstein et al. \cite{wasserstein2019moving}  suggested that we should not draw the conclusion that there is a statistically significant difference when $p<0.05$, or there is not a statistically significant difference when $p>0.05$.

Based on the above discussion and our experience,
our suggestions for the two questions are as follows.
For the statistical metrics,
as no single metric is perfect,
it is better to report a set of statistical metrics such as \emph{mean},
\emph{median},
the $p$ value, etc.
In addition to measuring whether fuzzer \emph{A} performs better than fuzzer \emph{B},
it is also important to measure how much fuzzer \emph{A} performs better than fuzzer \emph{B}.
To quantify the \emph{extent} of the difference between two fuzzers,
it is recommended to use the Vargha and Delaney $\hat{A}_{12}$ score \cite{vargha2000critique} to show the probability that fuzzer \emph{A} performs better than fuzzer \emph{B}.
For the number of repetitions, which is related to the selected statistical metrics.
For instance, 
the number of repetitions should be larger than 20 when using the Mann-Whitney U test~\cite{scipy-stats-mannwhitneyu}.
In addition,
it is necessary to conduct deeper research on these two questions in the future.

\textbf{Quality of Bugs.}
We define the quality of bugs from the perspective of evaluating fuzzers' performance.
That is, the quality of bugs should reflect not only the severity of bugs, but also the effectiveness of fuzzers in finding rare bugs.
A fuzzer which can find more high quality bugs should be considered as better.
Specifically, 
we measure the quality of a bug from two main aspects:
(1) \emph{whether a bug has a higher level of severity} and (2) \emph{whether a bug is harder to be found}.
For aspect (1),
we can leverage analysis tools to measure the severity of a bug.
For example,
\texttt{Exploitable}~\cite{exploitable} is a GDB extension that classifies Linux application bugs by severity.
Moreover,
we can map a bug with its corresponding CVE and assess its severity by the Common Vulnerability Score System (CVSS) score~\cite{CVSS} of the CVE.
For aspect (2),
a bug which is hard to be found usually has the following characteristics:
it can be found by few fuzzers or it is mapped with a small amount of crash samples.

\textbf{Speed of Finding Bugs.}
Finding bugs quickly and efficiently is important, especially when the time budget is limited.
We can measure the speed of finding bugs using the following two approaches.
First,
for all the bugs of a program,
we can draw the cumulative curve of the number of all the detected unique bugs within a pre-defined time.
A higher slope of the curve means a higher speed to find bugs,
which is a relatively qualitative metric.
Second,
for a specific bug,
we can record the \emph{time-to-exposure} (TTE) metric to measure the time that it is found by a fuzzer for the first time, which is a relatively quantitative metric.

\textbf{Stability of Finding Bugs.}
Stability is another important metric.
A fuzzer with higher stability is more reliable and practical.
We can quantify the stability of a fuzzer in the following manner.
First,
we can calculate the \emph{relative standard deviation} (RSD) of the number of the found unique bugs among all the repetitions.
Lower RSD means better stability.
Second,
for a specific bug,
we can calculate the number of times that a fuzzer can find it successfully.
Higher success rate represents that a fuzzer has better stability.

\textbf{Coverage.}
Coverage metrics are used to measure a fuzzer's capability of exploring paths,
which are also significant in quantifying the capability of a fuzzer.
As the vulnerable code usually takes a tiny fraction of the entire code,
only considering the number of bugs may not be able to distinguish the fuzzers' capability in exploring paths. 
Coverage metrics can be measured with different granularity levels such as function,
basic block,
edge and line coverage, etc.

\textbf{Overhead.}
The overhead metrics instead aim to quantify how many computing resources a fuzzer costs during the fuzzing process,
which are also important.
For instance, we may determine that a fuzzer performs well when we only consider how many bugs it finds, but the determination may be misleading if it costs  much more computing resources than others.
This metric is instructive for users who have limited computing resources or need to run multiple fuzzers in parallel.
The overhead of a fuzzer can be measured by the following concrete metrics:
\emph{CPU utilization},
\emph{memory consumption},
and \emph{the amount of disk read/write}, etc.

\section{Evaluations of the State-of-the-art Fuzzers}\label{general}

Leveraging \system,
we conduct extensive experiments on the state-of-the-art fuzzers,
and comprehensively compare them in terms of the six categories of performance metrics. 
Following the guidelines in \cite{klees2018evaluating},
we conduct fuzz testing for 24 hours, with 30 repetitions.

\begin{figure*}
	\centering
	\subfigure[Real-world programs]{
	\includegraphics[width=7in]{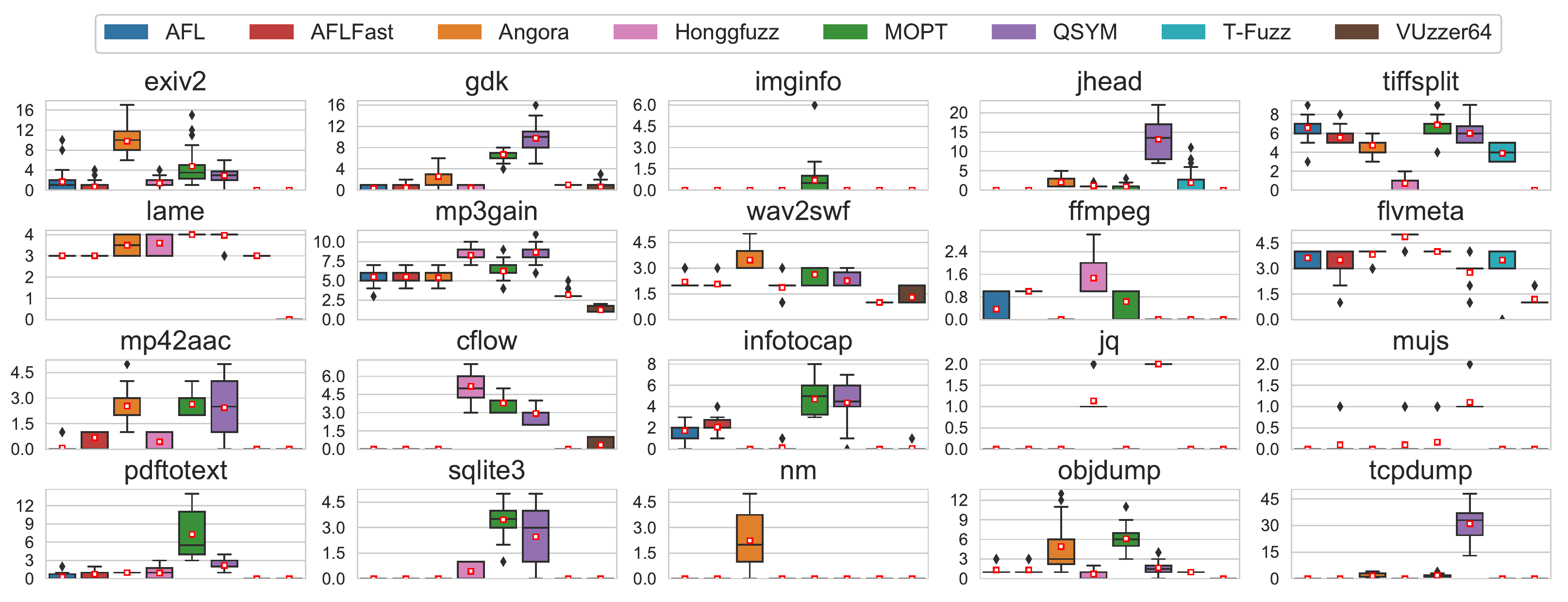}
	\label{fig-ge-bug-realworld}
	}\vspace{-3mm}
	\subfigure[LAVA-M programs]{
	\includegraphics[width=7in]{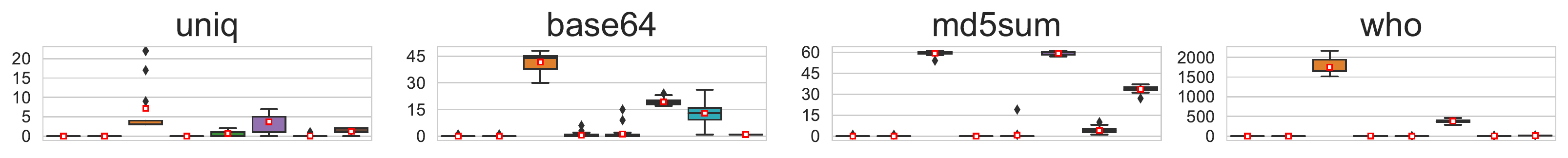}
	\label{fig-ge-bug-lava}
	}
\caption{The number of unique bugs detected by fuzzers.}
\end{figure*}

\subsection{Experiment Settings}\label{sub-expset}

\textbf{Fuzzers.}
In our evaluations,
we select eight state-of-the-art coverage-based fuzzers from \system, 
including AFL~\cite{AFL}, 
AFLFast~\cite{Pham2016Coverage}, 
Angora~\cite{Chen2018Angora}, 
Honggfuzz~\cite{honggfuzz}, 
MOPT~\cite{mopt},
QSYM~\cite{yun2018qsym}, 
T-Fuzz~\cite{T-Fuzz} and 
VUzzer64~\cite{Rawat2017VUzzer}.
The reasons for selecting these fuzzers are as follows.
First,
they are prominent fuzzers at the time of writing this paper.
AFL and Honggfuzz are proposed in industry which have been widely applied in practice.
The other six fuzzers are presented at top security conference in recent years,
which represent advanced fuzzing techniques in academia.
Second,
although there are other advanced fuzzers such as CollAFL \cite{CollAFL}, 
they are not open-source making them difficult to be evaluated.
Third,
the eight selected fuzzers have better scalability than others which can be used to test most of the programs.
In comparison,
other fuzzers such as QuickFuzz \cite{grieco2016quickfuzz} can only generate limited types of test cases\footnote{\scriptsize For instance, as QuickFuzz cannot generate .mp3 file, we cannot test QuickFuzz on program \texttt{mp3gain}.}.
Thus, they can only be tested on a limited number of benchmark programs.
Fourth,
as it is not appropriate to compare between fuzzers of different types,
here we only include coverage-based fuzzers to provide comparable evaluations.
For other fuzzers incorporated in \system, 
we mainly focus on testing their usability and making them available.
In addition, to make the evaluations fairer and more comparable, we make necessary modifications on several fuzzers.
For Angora, we change the input size limitation from 15 KB to 1 MB, to make it fairly comparable with the other fuzzers.
For VUzzer64, we fix the issues (\#10, \#11, \#12 and \#14) of its repository as these flaws are serious.
In addition, we modify the value of the variable \texttt{GENNUM}, which determines the number of generations, from 1,000 to 1,000,000 to make VUzzer64  fuzz for a longer time.
For T-Fuzz, we fix its ``naming" flaw.
All the above modifications are to make the evaluation fairer.
For the other fuzzers, we keep them the same as the original design.

\textbf{Programs.}
We leverage the 20 real-world programs (Table \ref{tab-bench}) provided by \system to evaluate the selected fuzzers.
In addition, we also use LAVA-M to explore the gap between the synthetic and the real-world programs.
Each benchmark program is compiled according to each fuzzer's requirements in instrumentation or compilation.
When validating the found bugs, the programs are compiled with ASan and GDB, which is same for all the fuzzers.

\textbf{Initial Seeds.}
Following common practice in fuzzing,
we utilize the same initial seeds for the same benchmark program.
For the \system benchmark programs,
we select the initial seeds by the following process.
First, we collect some seeds with the corresponding file format from the Internet.
Second, we exclude the seeds which do not satisfy a fuzzer's requirement (e.g., AFL requires the size of a seed be less than 1 MB).
Then, for the rest ones, we randomly select 100 seeds for each program.
For programs of LAVA-M,
we use the default seed set provided by LAVA-M.

\textbf{Environments.}
We conduct all the experiments on 5 servers with the same equipment: 20 Intel Xeon E5-2650 v4 CPU cores with 2.20 GHz, 64-bit Ubuntu 16.04 LTS.
For each fuzzer,
we assign one CPU core, 
2 GB RAM, and 1 GB swap space.
If a fuzzer cannot run successfully with 2 GB memory, 
we increase the memory limit to 8 GB, 
with the same resources allocated for all fuzzers in the same fuzzing experiment\footnote{\scriptsize We do not start all experiments at 8 GB due to resource limitations. As the same resources are allocated for all fuzzers in the same fuzzing experiment, the evaluations are fair.}. 
We run each fuzzing experiment in an independent Docker container.

Next we present the main evaluation results based on the six categories of performance metrics, and we provide more detailed evaluation results and datasets on the \system open-source platform \cite{unifuzz}.
Note that  T-Fuzz costs too much memory when fuzzing program \texttt{ffmpeg}, and VUzzer64 cannot test program \texttt{sqlite3} because it does not support input from \texttt{stdin}.
Thus, we do not include the results of the above cases.

\subsection{Quantity of Unique Bugs}\label{sub-bugnum}
The main objective of this subsection is to figure out
\emph{which fuzzer can find more unique bugs.}
As described in Section \ref{benchmark},
we leverage the output report produced by ASan \cite{AS} to extract the top three functions in the stack trace as a triple to de-duplicate bugs.
The bugs that have the different triples and vulnerability types are considered as unique.
For bugs that cannot be detected by ASan \cite{AS},
we further leverage the output report produced by other tools such as GDB \cite{gdb} as a supplement\footnote{\scriptsize When verifying the crashes found by T-Fuzz, if the crashes cannot make the original program crash, we then use CrashAnalyzer (provided by the developers of T-Fuzz) to generate new crashes and verify them. 
However, we find that whether using CrashAnalyzer has no impact on our experiment results. 
For the tested 20 programs, T-Fuzz does not generate transformed binaries for 17 programs (the reason is that T-Fuzz will not generate transformed binaries if there is no ``stuck" state), which means there is no need to use CrashAnalyzer. 
For the three programs (\texttt{jhead, flvmeta and wav2swf}), the CrashAnalyzer does not produce new crashes.},
as we find that there exist some bugs, e.g., \emph{float point exception} bugs which can be detected by GDB, while not ASan.

\textbf{Number of Unique Bugs.}
We visualize the number of unique bugs found by each fuzzer on the real-world programs of the \system benchmark and LAVA-M in 30 repetitions in Figure \ref{fig-ge-bug-realworld} and Figure \ref{fig-ge-bug-lava} respectively.
From these figures, we have the following observations.
(1) No fuzzer outperforms others on all the programs.
(2) For the 20 real-world programs,
QSYM performs the best on five programs (\texttt{gdk},
\texttt{jhead},
\texttt{lame},
\texttt{mujs},
\texttt{tcpdump}).
Angora performs the best on three programs 
(\texttt{exiv2}, \texttt{wav2swf}, \texttt{nm}).
Honggfuzz performs the best on three programs 
(\texttt{ffmpeg}, \texttt{flvmeta}, \texttt{cflow}).
MOPT performs the best on three programs (\texttt{imginfo}, \texttt{lame}, \texttt{pdftotext}).
AFL performs the best on program \texttt{tiffsplit}.
AFLFast, T-Fuzz and VUzzer64 fail to achieve the best performance on any target program.
(3) For the programs of LAVA-M,
Angora performs the best among the selected fuzzers,
while QSYM only achieves similar performance as Angora on program \texttt{md5sum}.

\begin{table*}[htb]
\centering
\footnotesize
\renewcommand\arraystretch{1.1}
\caption{The $p$ value and the $\hat{A}_{12}$ score of the number of unique bugs in 30 repetitions with AFL as the baseline.}
\label{tab-ge-bug-stat1}
\setlength{\tabcolsep}{4mm}{
\scalebox{0.7}{
\begin{tabular}{l|l|c|cc|cc|cc|cc|cc|cc|cc}
\hline	
&& \textbf{AFL} & \multicolumn{2}{c}{\textbf{AFLFast}}  & \multicolumn{2}{c}{\textbf{Angora}} & \multicolumn{2}{c}{\textbf{Honggfuzz}}& \multicolumn{2}{c}{\textbf{MOPT}}  & \multicolumn{2}{c}{\textbf{QSYM}} & \multicolumn{2}{c}{\textbf{T-Fuzz}} & \multicolumn{2}{c}{\textbf{VUzzer64}} \\
\hline
&&Avg & p-val & $\hat{A}_{12}$     & p-val   & $\hat{A}_{12}$   & p-val     & $\hat{A}_{12}$     & p-val   & $\hat{A}_{12}$  & p-val   & $\hat{A}_{12}$   & p-val    & $\hat{A}_{12}$ & p-val    & $\hat{A}_{12}$\\
\hline
\multicolumn{1}{c|}{\multirow{20}{*}{\rotatebox{90}{\textbf{Real-world Programs}}}}

&\texttt{exiv2} & 1.7 & 0.01 & 0.32 & \bm{$<0.01$} & \textbf{0.97} & 0.27 & 0.55 & \bm{$<0.01$} & \textbf{0.83} & \bm{$<0.01$} & \textbf{0.78} & $<0.01$ & 0.17 & $<0.01$ & 0.17 \\ 
&\texttt{gdk} & 0.3 & 0.11 & 0.58 & \bm{$<0.01$} & \textbf{0.92} & 0.21 & 0.55 & \bm{$<0.01$} & \textbf{1.0} & \bm{$<0.01$} & \textbf{1.0} & \bm{$<0.01$} & \textbf{0.85} & 0.18 & 0.56 \\ 
&\texttt{imginfo} & 0 & 1.0 & 0.5 & 1.0 & 0.5 & 1.0 & 0.5 & \bm{$<0.01$} & \textbf{0.75} & 1.0 & 0.5 & 1.0 & 0.5 & 1.0 & 0.5 \\ 
&\texttt{jhead} & 0 & 1.0 & 0.5 & \bm{$<0.01$} & \textbf{1.0} & \bm{$<0.01$} & \textbf{1.0} & \bm{$<0.01$} & \textbf{0.78} & \bm{$<0.01$} & \textbf{1.0} & \bm{$<0.01$} & 0.7 & 1.0 & 0.5 \\ 
&\texttt{tiffsplit} & 6.6 & $<0.01$ & 0.25 & $<0.01$ & 0.12 & $<0.01$ & 0 & 0.26 & 0.55 & 0.01 & 0.33 & $<0.01$ & 0.06 & $<0.01$ & 0 \\ 
&\texttt{lame} & 3 & 1.0 & 0.5 & \bm{$<0.01$} & \textbf{0.75} & \bm{$<0.01$} & \textbf{0.8} & \bm{$<0.01$} & \textbf{1.0} & \bm{$<0.01$} & \textbf{0.98} & 1.0 & 0.5 & $<0.01$ & 0 \\ 
&\texttt{mp3gain} & 5.5 & 0.28 & 0.46 & 0.13 & 0.42 & \bm{$<0.01$} & \textbf{1.0} & \bm{$<0.01$} & \textbf{0.71} & \bm{$<0.01$} & \textbf{0.98} & $<0.01$ & 0.03 & $<0.01$ & 0 \\ 
&\texttt{wav2swf} & 2.2 & 0.07 & 0.43 & \bm{$<0.01$} & \textbf{0.94} & $<0.01$ & 0.35 & \bm{$<0.01$} & \textbf{0.72} & 0.28 & 0.53 & $<0.01$ & 0 & $<0.01$ & 0.12 \\ 
&\texttt{ffmpeg} & 0.37 & \bm{$<0.01$} & \textbf{0.82} & $<0.01$ & 0.32 & \bm{$<0.01$} & \textbf{0.87} & \bm{$0.02$} & 0.63 & $<0.01$ & 0.32 & n.a. & n.a. & $<0.01$ & 0.32 \\ 
&\texttt{flvmeta} & 3.63 & 0.39 & 0.48 & \bm{$0.04$} & 0.6 & \bm{$<0.01$} & \textbf{0.96} & \bm{$<0.01$} & 0.68 & $<0.01$ & 0.21 & 0.37 & 0.52 & $<0.01$ & 0 \\ 
&\texttt{mp42aac} & 0.07 & \bm{$<0.01$} & \textbf{0.8} & \bm{$<0.01$} & \textbf{1.0} & \bm{$<0.01$} & 0.68 & \bm{$<0.01$} & \textbf{1.0} & \bm{$<0.01$} & \textbf{0.9} & 0.08 & 0.47 & 0.08 & 0.47 \\ 
&\texttt{cflow} & 0 & 1.0 & 0.5 & 1.0 & 0.5 & \bm{$<0.01$} & \textbf{1.0} & \bm{$<0.01$} & \textbf{1.0} & \bm{$<0.01$} & \textbf{1.0} & 1.0 & 0.5 & \bm{$<0.01$} & 0.67 \\ 
&\texttt{infotocap} & 1.73 & 0.08 & 0.6 & $<0.01$ & 0.03 & $<0.01$ & 0.06 & \bm{$<0.01$} & \textbf{0.97} & \bm{$<0.01$} & \textbf{0.85} & $<0.01$ & 0.03 & $<0.01$ & 0.05 \\ 
&\texttt{jq} & 0 & 1.0 & 0.5 & 1.0 & 0.5 & \bm{$<0.01$} & \textbf{1.0} & 1.0 & 0.5 & \bm{$<0.01$} & \textbf{1.0} & 1.0 & 0.5 & 1.0 & 0.5 \\ 
&\texttt{mujs} & 0 & \bm{$0.04$} & 0.55 & 1.0 & 0.5 & \bm{$0.04$} & 0.55 & \bm{$0.01$} & 0.58 & \bm{$<0.01$} & \textbf{1.0} & 1.0 & 0.5 & 1.0 & 0.5 \\ 
&\texttt{pdftotext} & 0.3 & \bm{$<0.01$} & 0.69 & \bm{$<0.01$} & \textbf{0.85} & \bm{$<0.01$} & \textbf{0.71} & \bm{$<0.01$} & \textbf{1.0} & \bm{$<0.01$} & \textbf{1.0} & $<0.01$ & 0.37 & $<0.01$ & 0.37 \\ 
&\texttt{sqlite3} & 0 & 1.0 & 0.5 & 1.0 & 0.5 & \bm{$<0.01$} & \textbf{0.72} & \bm{$<0.01$} & \textbf{1.0} & \bm{$<0.01$} & \textbf{0.93} & 1.0 & 0.5 & n.a. & n.a. \\ 
&\texttt{nm} & 0 & 1.0 & 0.5 & \bm{$<0.01$} & \textbf{0.93} & 1.0 & 0.5 & 1.0 & 0.5 & 1.0 & 0.5 & 1.0 & 0.5 & 1.0 & 0.5 \\ 
&\texttt{objdump} & 1.33 & 1.0 & 0.5 & \bm{$<0.01$} & \textbf{0.89} & $<0.01$ & 0.31 & \bm{$<0.01$} & \textbf{0.99} & 0.12 & 0.58 & 0.01 & 0.42 & $<0.01$ & 0 \\ 
&\texttt{tcpdump} & 0 & 1.0 & 0.5 & \bm{$<0.01$} & \textbf{0.88} & 1.0 & 0.5 & \bm{$<0.01$} & \textbf{1.0} & \bm{$<0.01$} & \textbf{1.0} & 1.0 & 0.5 & 1.0 & 0.5 \\

\hline
\multicolumn{1}{c|}{\multirow{4}{*}{\rotatebox{90}{\textbf{LAVA-M}}}} 
&\texttt{uniq} & 0 & 1.0 & 0.5 & \bm{$<0.01$} & \textbf{1.0} & 1.0 & 0.5 & \bm{$<0.01$} & \textbf{0.77} & \bm{$<0.01$} & \textbf{0.95} & 0.08 & 0.53 & \bm{$<0.01$} & \textbf{0.92} \\
&\texttt{base64} & 0.03 & 1.0 & 0.5 & \bm{$<0.01$} & \textbf{1.0} & \bm{$<0.01$} & 0.64 & \bm{$<0.01$} & 0.69 & \bm{$<0.01$} & \textbf{1.0} & \bm{$<0.01$} & \textbf{1.0} & \bm{$<0.01$} & \textbf{0.98} \\
&\texttt{md5sum} & 0.03 & 1.0 & 0.5 & \bm{$<0.01$} & \textbf{1.0} & 0.17 & 0.48 & 0.08 & 0.55 & \bm{$<0.01$} & \textbf{1.0} & \bm{$<0.01$} & \textbf{1.0} & \bm{$<0.01$} & \textbf{1.0} \\
&\texttt{who} & 0 & 1.0 & 0.5 & \bm{$<0.01$} & \textbf{1.0} & \bm{$<0.01$} & \textbf{1.0} & 0.08 & 0.53 & \bm{$<0.01$} & \textbf{1.0} & \bm{$<0.01$} & \textbf{1.0} & \bm{$<0.01$} & \textbf{1.0} \\
\hline
\end{tabular}}}
\end{table*}

\textbf{Statistical Results.}
Here we present the statistical results of the number of unique bugs found by the fuzzers in 30 repetitions.
Due to space limitation, we only present two statistical results: $p$ value and  $\hat{A}_{12}$ score. 
We defer the other statistical results such as mean and median values on the \system open-source platform \cite{unifuzz}.
The $p$ value aims to quantify whether there is a significant difference between two populations (corresponding to two fuzzers in our setting).
$\hat{A}_{12}$ score is used to measure the effect size (i.e., the probability that one fuzzer is better than the other according to all the repetitions).
Here, we use AFL as the baseline fuzzer following most previous works \cite{klees2018evaluating}.
Specifically, we leverage the Mann-Whitney U test to calculate the $p$ value, 
and we consider $p < 0.05 $ as an indicator that there exists a significant difference. 
For $\hat{A}_{12}$ score, we consider $\hat{A}_{12} \geq 0.71$ as an indicator that there is a large effect size \cite{vargha2000critique}.
Table \ref{tab-ge-bug-stat1} shows the $p$ value and  the $\hat{A}_{12}$ score of the number of unique bugs in 30 repetitions.
From Table \ref{tab-ge-bug-stat1}, 
we have the following observations.
(1)
None of the remaining seven fuzzers outperforms AFL significantly on all the real-world programs.
Nevertheless, there exists fuzzers (Angora, QSYM and VUzzer64) that outperform AFL significantly on all   four programs of LAVA-M.
(2)
Based on the results of $p$ value,
MOPT performs significantly better than AFL on 17 real-world programs, 
which is the most among the seven fuzzers.
QSYM, Angora and Honggfuzz perform significantly better than AFL on 13, 11 and 11 real-world programs respectively.
However,
AFLFast only performs significantly better than AFL on 4 real-world programs.
Even worse,
T-Fuzz and VUzzer64 do not outperform AFL significantly on any real-world program.
(3)
Considering the $\hat{A}_{12}$ score metric, the fuzzers have the similar performance compared with the $p$ value metric.
The fuzzers that have large effect size ($\hat{A}_{12} \geq 0.71$) all outperform significantly  ($p$ value < 0.05) than AFL, but not vice versa.
For instance, AFLFast outperforms significantly than AFL on \texttt{mujs} ($p$ = 0.04), but the effect size ($\hat{A}_{12} = 0.55$) is not large.

\subsection{The Quality of Bugs}

As explained in Section \ref{metrics},
we define bug quality based on their severity and the rareness.

\begin{table}[!t]
\footnotesize
\centering
\caption{The number of CVEs with high severity.}
\label{tab-cvss}
\setlength{\tabcolsep}{1.9mm}{
\scalebox{0.7}{
\begin{tabular}{l|cccccccc}
\hline
 & \textbf{AFL} & \textbf{AFLFast} & \textbf{Angora} & \textbf{Honggfuzz} & \textbf{MOPT} & \textbf{QSYM} & \textbf{T-Fuzz} & \textbf{VUzzer64}\\
\hline
\texttt{exiv2} & 1 & 1 & \textbf{4} & 2 & 3 & 2 & 0 & 0\\
\texttt{gdk} & 0 & 0 & 0 & 0 & 0 & 0 & 0 & 0\\
\texttt{imginfo} & 0 & 0 & 0 & 0 & \textbf{6} & 0 & 0 & 0\\
\texttt{jhead} & 0 & 0 & \textbf{1} & 0 & 0 & \textbf{1} & \textbf{1} & 0\\
\texttt{tiffsplit} & \textbf{3} & \textbf{3} & \textbf{3} & 1 & \textbf{3} & 2 & \textbf{3} & 0\\
\texttt{lame} & 1 & 1 & \textbf{2} & \textbf{2} & \textbf{2} & \textbf{2} & 1 & 0\\
\texttt{mp3gain} & 3 & 3 & 3 & \textbf{5} & \textbf{5} & \textbf{5} & 2 & 1\\
\texttt{wav2swf} & \textbf{1} & \textbf{1} & \textbf{1} & \textbf{1} & \textbf{1} & \textbf{1} & 0 & \textbf{1}\\
\texttt{ffmpeg} & 0 & 0 & 0 & \textbf{1} & 0 & 0 & n.a. & 0\\
\texttt{flvmeta} & 0 & 0 & 0 & 0 & 0 & 0 & 0 & 0\\
\texttt{mp42aac} & 0 & 0 & 0 & 0 & 1 & \textbf{3} & 0 & 0\\
\texttt{cflow} & 0 & 0 & 0 & 0 & 0 & 0 & 0 & 0\\
\texttt{infotocap} & 1 & 1 & 0 & 1 & \textbf{2} & \textbf{2} & 0 & 0\\
\texttt{jq} & 0 & 0 & 0 & \textbf{1} & 0 & \textbf{1} & 0 & 0\\
\texttt{mujs} & 0 & 0 & 0 & 0 & 0 & \textbf{1} & 0 & 0\\
\texttt{pdftotext} & 2 & 2 & 1 & 2 & \textbf{4} & 2 & 1 & 0\\
\texttt{sqlite3} & 0 & 0 & 0 & 0 & 0 & 0 & 0 & n.a.\\
\texttt{nm} & 0 & 0 & 0 & 0 & 0 & 0 & 0 & 0\\
\texttt{objdump} & 0 & 0 & 0 & 0 & 0 & 0 & 0 & 0\\
\texttt{tcpdump} & 0 & 0 & 4 & 0 & 3 & \textbf{57} & 0 & 0\\
\hline
\end{tabular}}}
\end{table}

\begin{table}[!t]
\footnotesize
\centering
\renewcommand\arraystretch{1}
\caption{The average number of unique \emph{EXPLOITABLE} bugs.}
\label{tab-exploitable}
\setlength{\tabcolsep}{1.9mm}{
\scalebox{0.7}{
\begin{tabular}{l|cccccccc}
\hline
& \textbf{AFL} & \textbf{AFLFast} & \textbf{Angora} & \textbf{Honggfuzz} & \textbf{MOPT}   & \textbf{QSYM}   & \textbf{T-Fuzz} & \textbf{VUzzer64 } \\
\hline
\texttt{exiv2} & 1.3 & 0.5 & \textbf{6.7} & 0.1 & 4.9 & 0.3 & 0 & 0 \\
\texttt{gdk} & 0.0 & 0.3 & 1.2 & 0 & \textbf{7.9} & 2.3 & 4.7 & 0.5 \\
\texttt{imginfo} & 0 & 0 & 0 & 0 & 0 & \textbf{0.03} & 0 & 0 \\
\texttt{jhead} & 0 & 0 & \textbf{0.2} & 0 & 0 & \textbf{0.2} & \textbf{0.2} & 0 \\
\texttt{tiffsplit} & 0.7 & \textbf{0.8} & 0.2 & 0 & \textbf{0.8} & 0 & 0.7 & 0 \\
\texttt{lame} & 1.0 & 1.0 & 5.8 & 3.4 & \textbf{9.2} & 3.1 & 1.0 & 0 \\
\texttt{mp3gain} & 0.1 & 0.1 & 0 & \textbf{2.0} & 0.8 & 0.8 & 0 & 0 \\
\texttt{wav2swf} & 3.0 & 3.1 & 3.0 & 0.1 & \textbf{10.0} & 3.0 & 0 & 0.3 \\
\texttt{ffmpeg} & 0 & 0 & 0 & \textbf{0.1} & 0 & 0 & n.a. & 0 \\
\texttt{flvmeta} & 0.2 & 0.3 & 0.1 & \textbf{0.6} & 0.5 & 0.1 & 0.1 & 0 \\
\texttt{mp42aac} & 0 & 0 & 0.0 & 0 & 0 & \textbf{0.5} & 0 & 0 \\
\texttt{cflow} & 0 & 0 & 0 & \textbf{0.8} & 0.2 & 0 & 0 & 0 \\
\texttt{infotocap} & 0 & 0 & 0 & 0 & \textbf{0.2} & 0 & 0 & 0 \\
\texttt{jq} & 0 & 0 & 0 & 0 & 0 & 0 & 0 & 0 \\
\texttt{mujs} & 0 & 0 & 0 & \textbf{0.1} & 0 & 0 & 0 & 0 \\
\texttt{pdftotext} & 0.3 & 0.7 & 1.0 & 0.7 & \textbf{7.2} & 5.0 & 0 & 0 \\
\texttt{sqlite3} & 0 & 0 & 0 & 0 & \textbf{3.1} & 0 & 0 & n.a. \\
\texttt{nm} & 0 & 0 & \textbf{4.8} & 0 & 0 & 0 & 0 & 0 \\
\texttt{objdump} & 0.3 & 0.4 & 1.2 & 0.5 & \textbf{3.2} & 0 & 0.2 & 0 \\
\texttt{tcpdump} & 0 & 0 & 0 & 0 & \textbf{0.3} & 0 & 0 & 0 \\
\hline
\end{tabular}}}
\end{table}

\subsubsection{Severity of Bugs}
The severity of bugs can be quantified by the CVE CVSS score \cite{CVSS} and the results of \texttt{Exploitable} \cite{exploitable}.

\textbf{CVE CVSS.}
CVSS \cite{CVSS} provides a numerical score for each CVE to quantify its severity.
A CVE is considered as highly severe when the score is greater than or equal to 7.0.
We leverage the \emph{CVE keywords database} and the matching method in Section \ref{sub-prog} to get the initial CVE matching results.
Then, we manually check the initial results to obtain the final matching results.
Further, we associate each CVE with the corresponding CVSS score.
During the CVE matching process, we also find six new CVEs:
\emph{CVE-2019-17450, CVE-2019-17451, CVE-2019-17594, CVE-2019-17595, CVE-2019-18359 and CVE-2019-19035}.
Table \ref{tab-cvss} shows the number of CVEs with high severity found by the fuzzers, and we provide more detailed information of the found CVEs on the \system open-source platform  \cite{unifuzz}, including the concrete CVSS score and the vulnerability type.
As presented in Table \ref{tab-cvss},
the fuzzers have preference on specific programs in discovering highly severe CVEs.
For instance,
QSYM discovers 57 highly severe CVEs on \texttt{tcpdump},
while Honggfuzz cannot discover any one.
However,
for \texttt{ffmpeg},
Honggfuzz can discover one highly severe CVE,
while the remaining fuzzers (including QSYM) cannot find any one.
Moreover, it is interesting to note that AFL and AFLFast are comparable with respect to the number of discovered highly severe CVEs on each program.

\textbf{Results of \texttt{Exploitable}.}
\texttt{Exploitable} \cite{exploitable} is a GDB extension that uses a heuristic algorithm to assess the exploitability of a crash,
which can be classified into four categories: \emph{EXPLOITABLE},
\emph{PROBABLY\_EXPLOITABLE},
\emph{PROBABLY\_NOT\_EXPLOITABLE} and \emph{UNKNOWN}.
Specifically,
we de-duplicate the number of bugs of each category by the hash value produced by \texttt{Exploitable}.
Table~\ref{tab-exploitable} presents the  of unique bugs that are classified as \emph{EXPLOITABLE}.
As presented in Table \ref{tab-exploitable},
MOPT outperforms the other fuzzers on 9 programs in detecting \emph{EXPLOITABLE} bugs.
Angora, Honggfuzz and QSYM achieve the best performance on 3, 5 and 3 programs, respectively.
Nevertheless, VUzzer64 does not perform well as it can only detect \emph{EXPLOITABLE} bugs on 2 programs.

\begin{table}[tb]
\footnotesize
\centering
\renewcommand\arraystretch{1.1}
\setlength{\tabcolsep}{0.5pt}
\caption{The number of discovered unique \emph{rare bugs}.}
\label{tab-rareness}
\setlength{\tabcolsep}{1.9mm}{
\scalebox{0.7}{
\begin{tabular}{l|cccccccc}
\hline
& \textbf{AFL} & \textbf{AFLFast} & \textbf{Angora} & \textbf{Honggfuzz} & \textbf{MOPT} &\textbf{QSYM}  & \textbf{T-Fuzz} & \textbf{VUzzer64} \\
\hline
\texttt{exiv2} & 8 & 1 & 17 & 0 & \textbf{22} & 0 & 0 & 0\\
\texttt{gdk} & 0 & 0 & 2 & 0 & 1 & \textbf{13} & 0 & 1\\
\texttt{imginfo} & 0 & 0 & 0 & 0 & \textbf{7} & 0 & 0 & 0\\
\texttt{jhead} & 0 & 0 & 1 & 0 & 0 & \textbf{15} & 2 & 0\\
\texttt{tiffsplit} & 0 & 0 & 0 & 0 & \textbf{3} & 2 & 0 & 0\\
\texttt{lame} & 0 & 0 & 0 & 0 & 0 & 0 & 0 & 0\\
\texttt{mp3gain} & 0 & 0 & 0 & \textbf{1} & 0 & 0 & 0 & 0\\
\texttt{wav2swf} & 0 & 0 & \textbf{2} & 0 & 0 & 0 & 0 & 0\\
\texttt{ffmpeg} & 0 & 0 & 0 & \textbf{3} & 0 & 0 & n.a. & 0\\
\texttt{flvmeta} & 0 & 0 & 0 & \textbf{1} & 0 & 0 & 0 & 0\\
\texttt{mp42aac} & 0 & 0 & 2 & 0 & \textbf{8} & 7 & 0 & 0\\
\texttt{cflow} & 0 & 0 & 0 & \textbf{2} & \textbf{2} & 0 & 0 & 0\\
\texttt{infotocap} & 0 & 0 & 0 & 0 & 3 & \textbf{4} & 0 & 0\\
\texttt{jq} & 0 & 0 & 0 & 0 & 0 & 0 & 0 & 0\\
\texttt{mujs} & 0 & 0 & 0 & \textbf{2} & 0 & \textbf{2} & 0 & 0\\
\texttt{pdftotext} & 0 & 0 & 0 & 0 & \textbf{35} & 7 & 0 & 0\\
\texttt{sqlite3} & 0 & 0 & 0 & 0 & 1 & \textbf{3} & 0 & n.a.\\
\texttt{nm} & 0 & 0 & \textbf{25} & 0 & 0 & 0 & 0 & 0\\
\texttt{objdump} & 0 & 1 & \textbf{6} & 0 & 4 & 5 & 0 & 0\\
\texttt{tcpdump} & 0 & 0 & 1 & 0 & 4 & \textbf{204} & 0 & 0\\
\hline
Total & 8 & 2 & 56 & 9 & 90 & \textbf{262} & 2 & 1\\
\hline
\end{tabular}}}
\end{table}

\subsubsection{Rareness of Bugs}
It is intuitive that a bug that can be found by fewer fuzzers is relatively harder to be found (e.g., is located in deeper path or has more complicated path constraints).
Here,
we call a bug that can be found by only one fuzzer a \emph{rare bug}\footnote{\scriptsize The value of this metric is not an absolute value such as the number of unique bugs, while providing a relative measure that depends on the compared fuzzers. }.
Correspondingly, 
a fuzzer that can find more unique \emph{rare bugs} is relatively more powerful.
Table \ref{tab-rareness} shows the number of unique \emph{rare bugs} discovered by the evaluated fuzzers.
For all the real-world programs,
QSYM achieves the best performance by discovering 262 unique \emph{rare bugs}.
MOPT achieves the second best performance and discovers 90 unique \emph{rare bugs}.
Angora finds 56 unique \emph{rare bugs} in total.
Nevertheless,
AFLFast only detects \emph{rare bugs} on two programs.
AFL, T-Fuzz and VUzzer64 only detect \emph{rare bugs} on one program.
It is worth noting that fuzzers also have preference on target programs in discovering \emph{rare bugs}. 
For instance, 
QSYM discovers 204 unique \emph{rare bugs} on \texttt{tcpdump},
while in comparison Angora only discovers one \emph{rare bug} and the other fuzzers fail to find any \emph{rare bug}.
For \texttt{nm},
Angora can discover 25 unique \emph{rare bugs},
while the remaining fuzzers including QSYM fail to discover any \emph{rare bug}.

\begin{figure*}
	\centering
	\includegraphics[width=7in]{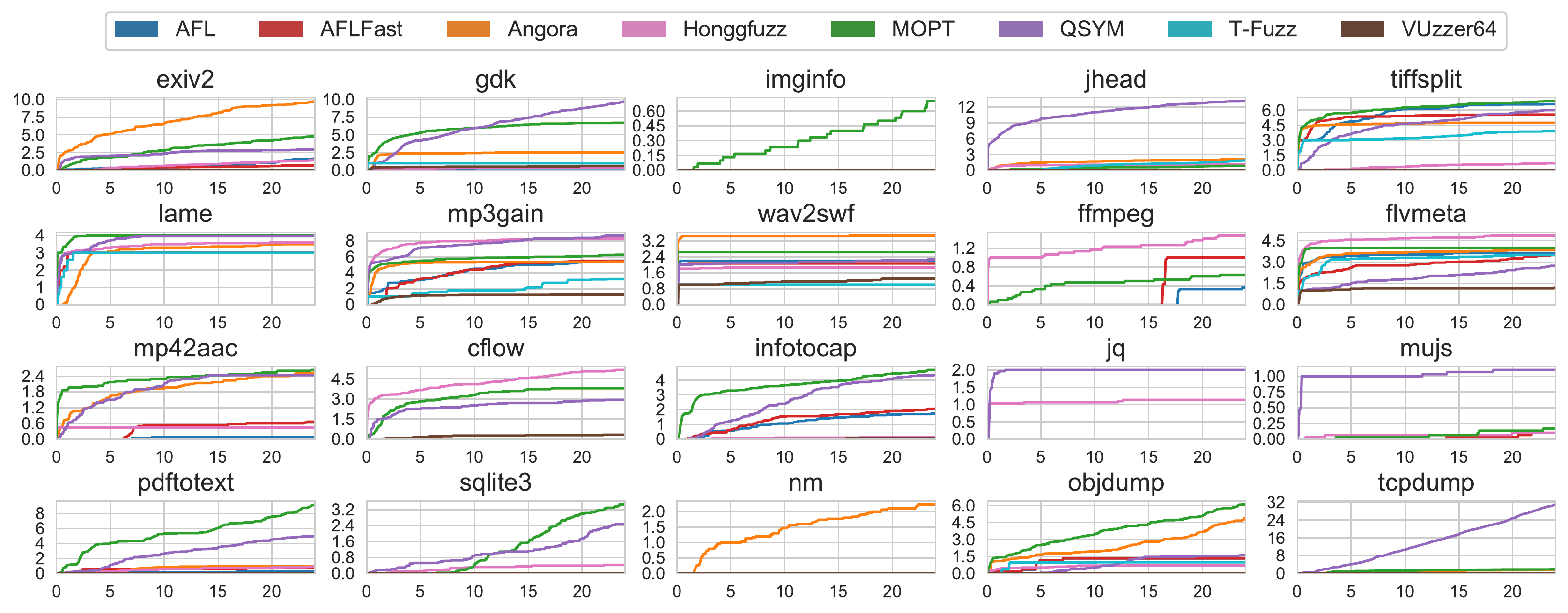}
	\caption{The average number of unique bugs found over time in 30 repetitions.}
	\label{fig-speed}
\end{figure*}

\subsection{Speed of Finding Bugs}
Figure~\ref{fig-speed} presents the average number of unique bugs found over time in 30 repetitions,
where we can see the fuzzers' speed of finding bugs.
First, 
one intuitive observation is that no fuzzer wins the others on all the programs on this metric.
Second, 
 the comparisons among fuzzers' performance may get reverse over time.
For instance,
MOPT finds less unique bugs than QSYM on program \texttt{sqlite3} in the early time, but it finds more unique bugs than QSYM after 10 hours.
Third,
although some fuzzers find the similar number of unique bugs,
their speeds of finding bugs are different.
For instance,
Angora, MOPT and QSYM find a similar number of unique bugs on program \texttt{mp42aac} (2.5, 2.6 and 2.4 unique bugs in average, respectively) within 24 hours,
while MOPT finds the bugs more quickly than Angora and QSYM.
This observation also indicates the importance of the speed metric, as only leveraging the number of unique bugs may overlook the difference of fuzzers in speed.

\subsection{Stability of Finding Bugs}
\begin{figure}[!t]
	\centering
	\includegraphics[width=3.5in]{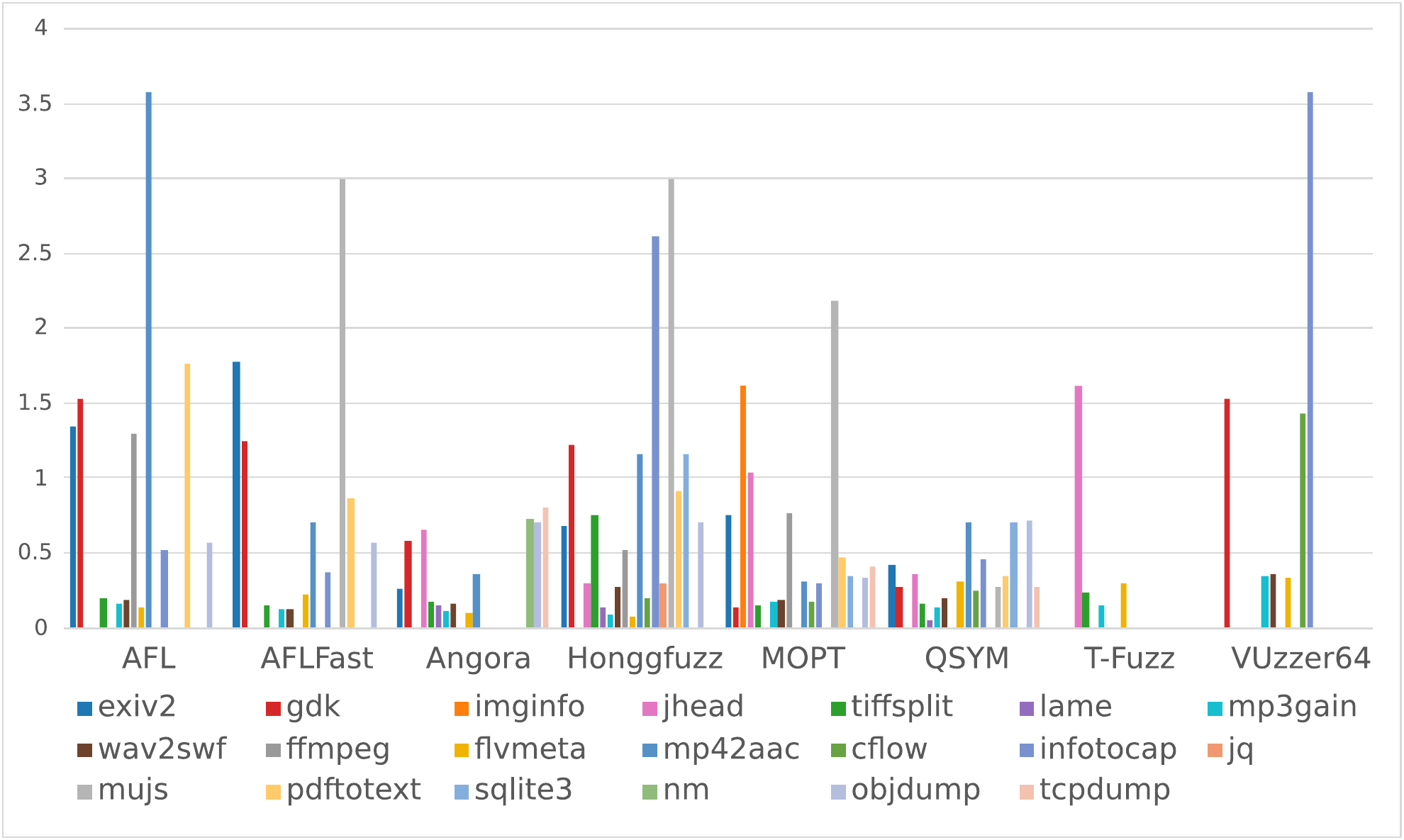}
	\caption{The RSD of the number of unique bugs.}
	\label{fig-bug-rsd}
\end{figure}

Figure \ref{fig-bug-rsd} presents the \emph{relative standard deviation} (RSD) of the number of unique bugs in all the 30 repetitions,
where a lower RSD represents better stability of a fuzzer.
As depicted in Figure \ref{fig-bug-rsd},
first,
all the fuzzers are not always
stable in finding bugs,
which reflects the randomness of fuzzing and the importance of repetitions.
Second,
among the seven fuzzers,
Angora and T-Fuzz achieve lower RSD,
while AFL and Honggfuzz achieve higher RSD.
Third,
the stability of a fuzzer varies with different programs.
For instance,
AFL has better stability on several programs such as  \texttt{tiffsplit}, \texttt{mp3gain},
as compared to that on \texttt{mp42aac}.
It needs to be noted that the stability of finding bugs metrics are auxiliary to the quantity of unique bugs metrics, as finding more bugs are more important than finding less bugs stably.

\subsection{Coverage}
\begin{figure}[htb]
    \centering
    \includegraphics[width=3.5in,height=1.5in]{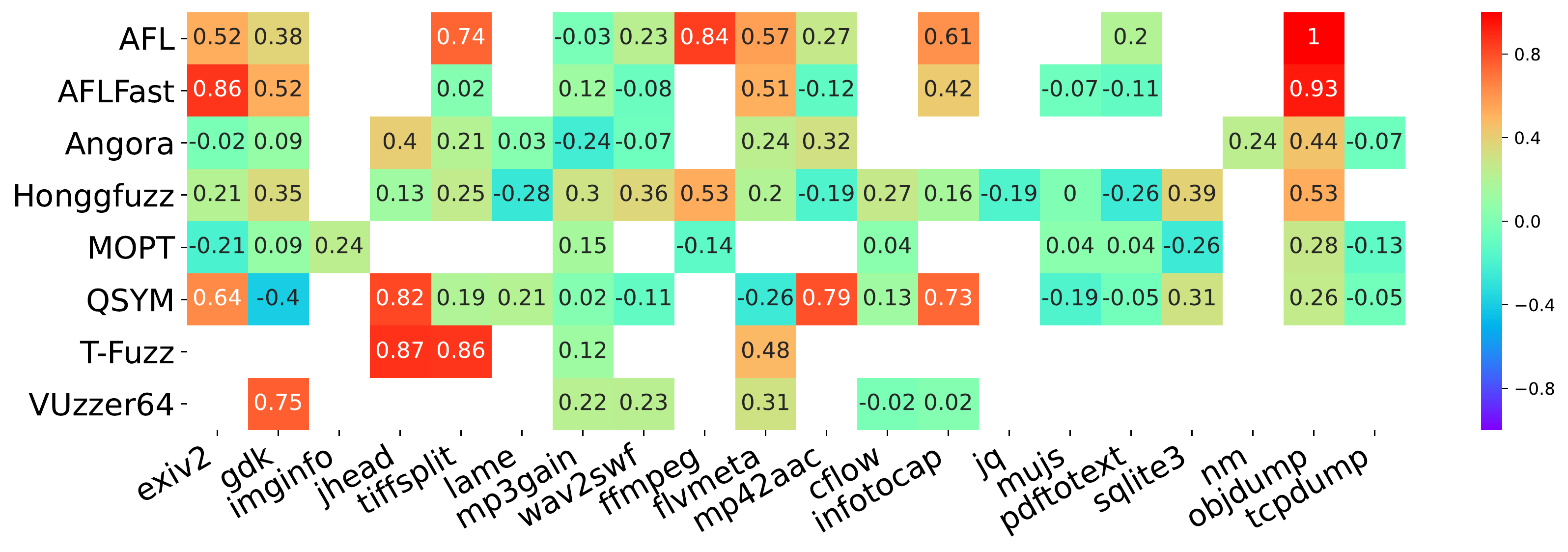}
    \caption{The Spearman's correlation coefficient $r_{s}$ between the number of unique bugs and line coverage. }
    \label{fig-rs}
\end{figure}

\begin{figure*}[htb]
	\centering
	\includegraphics[width=7in]{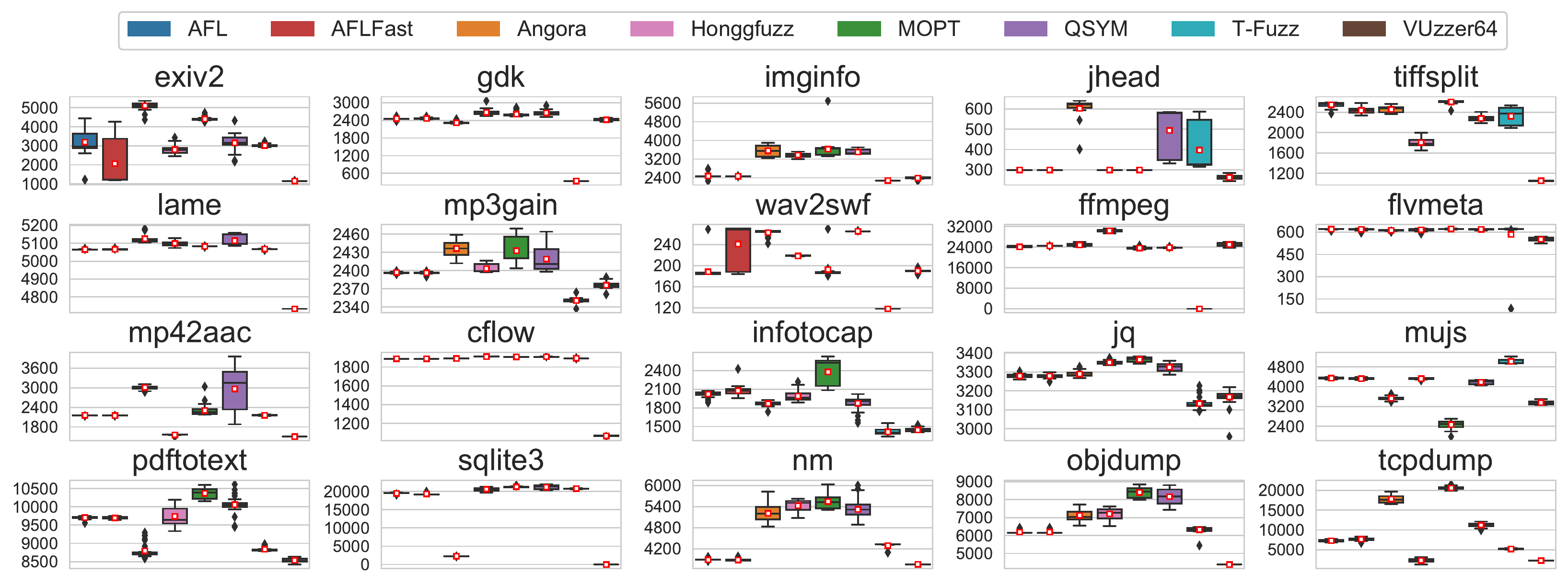}
	\caption{Line coverage on the real-world programs.}
	\label{fig-ge-line-coverage}
\end{figure*}

Existing fuzzers track the coverage of a target program with different manners and granularities.
For instance,
AFL~\cite{AFL} leverages compile-time instrumentation and bitmap to track edge coverage.
Honggfuzz~\cite{honggfuzz} leverages SanitizerCoverage instrumentation~\cite{sanitizer-coverage} to track basic block coverage.
In order to fairly compare these fuzzers' capability of finding paths,
it is necessary to design a uniform method (with the same instrumentation method and under the same granularity) to track the coverage for different fuzzers.
One intuitive method is to save all the test cases executed by the fuzzers,
then calculate their coverage with the same instrumented binary program.
Nevertheless,
this method is impractical as the number of executed test cases is tremendous.
To strike a balance between precision and efficiency,
we develop an efficient method to track the coverage by only considering the test cases that improve the coverage.
Specifically,
we save all the test cases that increase the coverage during the fuzzing process,
then we leverage afl-cov~\cite{afl-cov} to calculate the line coverage of each program with the saved test cases.
Figure~\ref{fig-ge-line-coverage} shows the results of line coverage,
from which we observe that no fuzzer stably achieves higher coverage than the others.
By comparing Figure~\ref{fig-ge-bug-realworld} and Figure~\ref{fig-ge-line-coverage},
we observe that \emph{higher coverage does not necessarily mean more unique bugs}.
For instance,
MOPT achieves the highest coverage on \texttt{tcpdump} among all the fuzzers
while QSYM discovers the most unique bugs on \texttt{tcpdump}.
To further explore the relationship between the number of unique bugs and line coverage,
we calculate the \emph{Spearman correlation coefficient} $r_{s}$ between them, which is a non-parametric measure of correlation between two variables and $r_{s} \in [-1,+1]$.
A positive $r_{s}$ means that the two variables are positively correlated and vice versa.
Figure~\ref{fig-rs} presents the value of $r_{s}$ between the number of unique bugs and line coverage,
where we observe that most of them are less than $0.60$, 
which means that the correlation between the number of unique bugs and the line coverage is not strong.

\begin{table}[htb]
\centering
\footnotesize
\renewcommand\arraystretch{1.15}
\setlength{\abovecaptionskip}{3pt} 
\setlength{\belowcaptionskip}{11pt}
\setlength{\tabcolsep}{0.9pt}
\caption{The memory consumption (MB) of each fuzzer.}
\label{tab-memory}
\scalebox{0.65}{
\begin{tabular}{l|cc|cc|cc|cc|cc|cc|cc|cc}
\hline
& \multicolumn{2}{c}{\textbf{AFL}}  & \multicolumn{2}{c}{\textbf{AFLFast}} & \multicolumn{2}{c}{\textbf{Angora}} & \multicolumn{2}{c}{\textbf{Honggfuzz}} & \multicolumn{2}{c}{\textbf{MOPT}} & \multicolumn{2}{c}{\textbf{QSYM}}   & \multicolumn{2}{c}{\textbf{T-Fuzz}}  & \multicolumn{2}{c}{\textbf{VUzzer64}}\\
\hline
& avg  & max & avg  & max & avg  & max & avg  & max  & avg  & max  & avg  & max  & avg  & max  & avg  & max\\
\hline
\texttt{exiv2} & 11 & 25 & 13 & 37 & 23 & 918 & 423 & 1,989 & 24 & 41 & 209 & 993 & 3,319 & 4,051 & 96 & 1,139 \\
\texttt{gdk} & 8 & 454 & 8 & 14 & 22 & 575 & 350 & 1,831 & 13 & 76 & 91 & 401 & 442 & 454 & 104 & 2048 \\
\texttt{imginfo} & 5 & 10 & 5 & 51 & 23 & 774 & 140 & 1,990 & 18 & 200 & 119 & 1,979 & 919 & 925 & 42 & 2,048 \\
\texttt{jhead} & 7 & 12 & 7 & 20 & 13 & 28 & 60 & 77 & 8 & 13 & 79 & 188 & 265 & 313 & 18 & 30 \\
\texttt{tiffsplit} & 16 & 213 & 14 & 73 & 22 & 55 & 100 & 1,930 & 18 & 654 & 42 & 97 & 567 & 765 & 122 & 474 \\
\texttt{lame} & 13 & 17 & 13 & 22 & 1,705 & 2,047 & 53 & 76 & 20 & 29 & 81 & 148 & 758 & 1,038 & 56 & 450 \\
\texttt{mp3gain} & 9 & 12 & 12 & 16 & 34 & 46 & 34 & 48 & 17 & 23 & 104 & 276 & 346 & 354 & 109 & 468 \\
\texttt{wav2swf} & 40 & 53 & 20 & 82 & 76 & 415 & 444 & 4,087 & 16 & 94 & 134 & 471 & 598 & 682 & 114 & 2,035 \\
\texttt{ffmpeg} & 17 & 596 & 19 & 502 & 27 & 246 & 734 & 5,533 & 77 & 1,254 & 212 & 1,780 & n.a. & n.a. & 849 & 8,195 \\
\texttt{flvmeta} & 9 & 14 & 9 & 14 & 19 & 27 & 24 & 50 & 12 & 15 & 598 & 1,873 & 470 & 694 & 154 & 318 \\
\texttt{mp42aac} & 8 & 23 & 9 & 15 & 58 & 532 & 60 & 1745 & 18 & 176 & 222 & 2,826 & 1,155 & 1,194 & 112 & 670 \\
\texttt{cflow} & 6 & 7 & 7 & 8 & 1,133 & 2,023 & 38 & 60 & 23 & 35 & 125 & 597 & 479 & 489 & 261 & 1,978 \\
\texttt{infotocap} & 14 & 23 & 15 & 40 & 24 & 27 & 316 & 428 & 24 & 38 & 496 & 1,361 & 597 & 606 & 184 & 636 \\
\texttt{jq} & 9 & 11 & 9 & 12 & 13 & 15 & 50 & 72 & 13 & 16 & 78 & 113 & 619 & 783 & 49 & 392 \\
\texttt{mujs} & 17 & 45 & 16 & 30 & 552 & 1,533 & 52 & 88 & 23 & 44 & 113 & 2,013 & 578 & 729 & 56 & 1,623 \\
\texttt{pdftotext} & 27 & 76 & 27 & 40 & 4,857 & 7,861 & 161 & 1,967 & 92 & 149 & 139 & 1,786 & 2,050 & 2,055 & 396 & 1,190 \\
\texttt{sqlite3} & 240 & 595 & 205 & 2,031 & 1,833 & 2,042 & 199 & 1,249 & 453 & 1,560 & 780 & 1,790 & 1,580 & 2,095 & n.a. & n.a. \\
\texttt{nm} & 8 & 34 & 8 & 26 & 102 & 1,350 & 279 & 2,046 & 35 & 50 & 78 & 350 & 1,739 & 2,265 & 57 & 460 \\
\texttt{objdump} & 13 & 171 & 13 & 70 & 108 & 574 & 495 & 2,048 & 49 & 1,953 & 137 & 2,698 & 2,625 & 3,472 & 849 & 1,368 \\
\texttt{tcpdump} & 15 & 38 & 16 & 38 & 264 & 607 & 330 & 2,040 & 83 & 107 & 160 & 350 & 1,464 & 2,296 & 119 & 322 \\
\hline
Avg & 24.6 & 121.5 & 22.2 & 157.1 & 545.4 & 1,084.8 & 217.1 & 1,467.7 & 51.8 & 326.4 & 199.8 & 1,104.5 & 1,082.6 & 1,329.5 & 197.2 & 1,360.2 \\
\hline
\end{tabular}}
\end{table}

\subsection{Overhead}\label{sub-overhead}
Table \ref{tab-memory} shows the average and maximum memory consumption of each fuzzer, 
where we obtain the following observations.
(1) From a holistic aspect, 
AFL, AFLFast and MOPT consume less memory during fuzzing than the other fuzzers, with average memory consumption 24.6 MB, 22.2 MB and 51.8 MB respectively.
Nevertheless,
T-Fuzz consumes 1,082 MB memory during fuzzing,
which is almost 50 times more than that of AFLFast and is the most among the fuzzers.
One possible reason is that T-Fuzz leverages Angr~\cite{shoshitaishvili2016state} to get the \emph{Control Flow Graph (CFG)} of the programs, which may take much memory.
(2)
When fuzzing the same program,
the memory consumption of different fuzzers varies significantly.
For example,
when fuzzing program \texttt{exiv2},
AFL uses no more than 25 MB memory,
while in comparison,
T-Fuzz uses about 4 GB memory.
(3)
For the same fuzzer,
its memory consumption on various programs also differ greatly.
For instance,
Angora uses more than 7 GB memory when testing \texttt{pdftotext} while its memory consumption on the other programs is less than 2 GB.

\section{Further Analysis}\label{casestudy}
Here we conduct evaluations to investigate the previously overlooked factors that may significantly affect a fuzzer's performance, including \emph{instrumentation methods} and \emph{crash analysis tools}.

\subsection{Instrumentation Methods}
Fuzzers may implement instrumentations in different manners,
which leads to diverse characteristics in the compiled binaries.
For instance,
AFL and Angora implement compile-time instrumentation by writing a wrapper of a compiler (e.g., \texttt{afl-clang},  \texttt{angora-clang}),
while VUzzer leverages Intel PIN \cite{Luk2005Pin} to implement binary instrumentation.
Therefore, 
a natural question is: 
\emph {whether different instrumentation methods affect fuzzing evaluation?}
We raise this question based on the following observations.
For programs such as  \texttt{infotocap}, 
certain crash samples can only make the AFL-instrumented binary crash rather than Angora-instrumented binary.
By analyzing these bugs of \texttt{infotocap},
we find that they are related to the compilation methods.
In this scenario,
the failure for Angora to discover these crash samples is due to its instrumentation method rather than its capability of discovering bugs. 
However, 
if we overlook the employed different instrumentation methods,
we may draw a potentially unfair conclusion that AFL is better than Angora with respect to the detected bugs in this scenario.

Here,
we provide an example to show that the compilation methods can impact the bugs.
Figure \ref{fig-heap} shows a C/C++ code snippet,
where there is a heap buffer overflow vulnerability in line 4.
However,
certain compile-time optimization may skip the erroneous assignment (\texttt{x[i]=0} in line 4) and treat the whole assignment statement as a constant zero to print.
We compile this code with different compilers (\texttt{gcc} and \texttt{clang}) and with different optimization levels (O0  - O3).
Then, 
we find that the heap buffer overflow vulnerability cannot be triggered when using \texttt{clang} with optimization level O1 - O3, 
but can be triggered when using \texttt{gcc} with optimization level O0 - O3 and \texttt{clang} with optimization level O0.
As it is difficult to require all the fuzzers to use the same instrumentation method in practice,
the difference caused by different compilation (instrumentation) methods cannot be avoided. 
Therefore, 
a potentially better solution is using \emph{cross validation} when analyzing the crash samples, i.e., re-execute the crash samples with different complied binaries to check if they can only cause parts of binaries  to crash.

\begin{figure}[!t]
	\begin{lstlisting}[
	language = C++, 
	numbers=left, 
	numberstyle=\scriptsize,
	numbersep=-5pt,
	basicstyle=\scriptsize,
	keywordstyle=\color{blue!70},
	commentstyle=\color{red!50!green!50!blue!50},
	frame=shadowbox,
	showtabs=false,
	tabsize=1,
	breaklines=true,
	rulesepcolor=\color{red!20!green!20!blue!20}]
    	char* x=malloc(1);
	    for(int i=0;i<1000000;i++){
	        /* x[i]=0 is a heap buffer overflow */
	        printf("[%d]=%c\n", i, x[i]=0);
	    }
	\end{lstlisting}
	\caption{An example of heap buffer overflow vulnerability.}
	\label{fig-heap}
\end{figure}

\subsection{Crash Analysis Tools}
Different tools have been proposed to analyze what bugs can be triggered by crash samples such as ASan \cite{AS} and GDB \cite{gdb}.
During our evaluations,
interestingly,
we find that using different tools to analyze crash samples can lead to different results,
e.g., different numbers of discovered unique bugs.
To further examine this,
we use ASan and GDB to analyze the crash samples collected from the experiments in Section \ref{general}.
If a bug can be discovered by executing the ASan- (resp., GDB-) compiled binary with a crash sample,
we consider the corresponding crash sample as validated with ASan (resp., GDB).
To show the influence of different tools in analyzing crash samples, 
we list the number of crash samples that can be validated by different tools in Table \ref{tab-validate}. 
For the collected 329,857 crash samples,
only 61.1\% of them can be validated by both ASan and GDB.
14.5\% of them can only be validated by GDB and 12.2\% of them can only be validated by ASan.
Moreover,
neither tool can validate the remaining 12.2\% crash samples.
It is a bit surprising to see that ASan, 
as a widely adopted analysis tool, 
only validates 73.3\% (12.2\%+61.1\%) of these crash samples.

\begin{figure}[htb]
	\centering
	\includegraphics[width=3in]{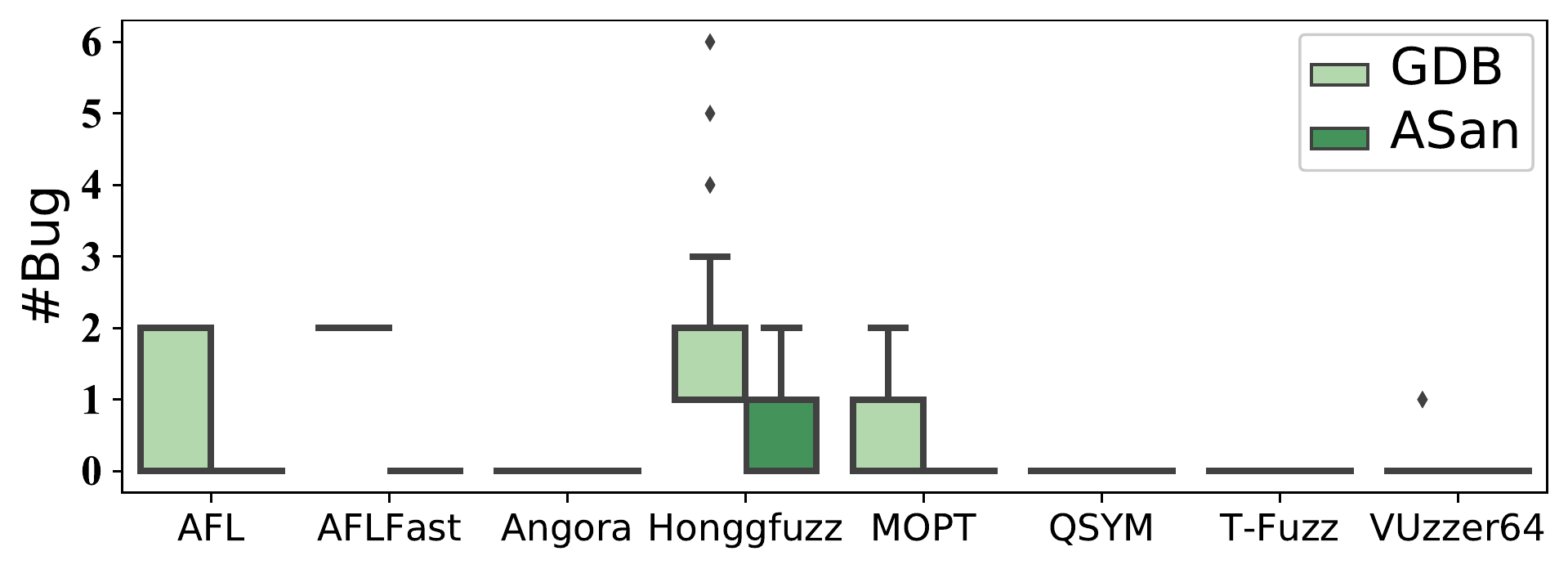}
	\caption{The number of unique bugs discovered on \texttt{ffmpeg} with GDB  and ASan.}
	\label{fig-gdb-asan}
\end{figure}

Using one analysis tool singly may limit the number of detected bugs,
which may further fail to provide comprehensive evaluations on fuzzers. 
For instance,
during our fuzzing experiments in Section \ref{general},
we find some crash samples that can trigger 
\emph{ float point exception} bugs on \texttt{ffmpeg}.
However, 
we cannot discover the float point exception bugs by executing ASan-compiled binary with these crash samples,
while GDB can discover them.
Figure \ref{fig-gdb-asan} shows the number of unique bugs discovered by the fuzzers on \texttt{ffmpeg} with ASan and GDB.
As shown in Figure \ref{fig-gdb-asan}, the evaluation results are different when using different analysis tools.
When using ASan,
only Honggfuzz can discover bugs,
while using GDB,
AFL and AFLFast can also discover bugs.
Therefore,
it would be better to combine more tools together to analyze crash samples instead of relying on a single tool that may neglect some bugs.
In our evaluation (Section \ref{general}),
we use ASan as the main tool to detect bugs,
while adopting GDB as a supplement.

\begin{table}[htb]
	\centering
	\caption{Validated crash samples by different tools.}
	\label{tab-validate}
	\setlength{\tabcolsep}{5mm}{
		\scalebox{0.8}{
			\begin{tabular}{l|cc}
				\hline
				\textbf{Bug Type} & \textbf{Number} & \textbf{Rate}     \\
				\hline
				Neither ASAN or GDB can validate  & 40,122  & 12.2\% \\
				Only GDB can validate     & 47,910  & 14.5\% \\
				Only ASAN can validate  & 40,267  & 12.2\%  \\
				Both ASAN and GDB can validate & 201,558 & 61.1\%  \\
				\hline
				Total & 329,857 & 100\% \\	
				\hline
	\end{tabular}}}
\end{table}

\section{Discussion}\label{discussion}

We discuss the issues in  the current fuzzing research field along with our work in this paper as follows.

\subsection{The Usability of Fuzzers}

A fuzzer with good usability can facilitate its application in practice.
Nevertheless, based on our evaluations, we find that the usability issues of fuzzers, especially the academic fuzzers, are more serious than we thought.
These usability issues include: having (serious) flaws in implementation, failing to be reproduced, etc.
Even worse, some of the fuzzers that have these issues are published at premier conferences in recent years.
In this paper, we test the usability of 35 fuzzers, make them available on the \system platform, and conduct extensive evaluations on eight of them.
We hope that this work can facilitate further research on improving the usability and the performance of fuzzers.

How to conduct a comprehensive measurement on the usability of fuzzers is an interesting and significant research topic.
Nevertheless, the usability of fuzzers is a relatively subjective topic, and it is affected by many factors.
First, the usability of a fuzzer highly depends on the domain knowledge of the users.
A fuzzing expert may easily use a fuzzer even without the documentation, while a beginner may feel hard to use a fuzzer with poor documentation.
Second,  there are many factors that can affect the usability of a fuzzer including documentation style, the issues of dependent libraries and tools, the issues of its implementation,  the robustness in fuzzing process, etc.
For the future research on this problem, we provide the following feasible ways as suggestions.
(1) Check the correctness and completeness of a fuzzer's documentation (e.g., whether there exists inconsistency between the documentation and the implementation). 
It is also an interesting and meaningful research topic on providing guidance or standards on writing the documentation.
(2) Test whether a fuzzer can be successfully installed and pass author-provided tests.
(3) Test the robustness of a fuzzer during the fuzzing process, and observe whether it has abnormal behaviors (e.g., whether the fuzzer itself crashes during fuzzing).
(4) Test whether a fuzzer can reproduce the experimental results as it reported in its paper.

\subsection{Fuzzing Experiments}
Conducting correct fuzzing experiments is the base of the appropriate evaluations.
Klees et al. \cite{klees2018evaluating} proposed several guidelines in fuzzing evaluation such as multiple repetitions, using different seed sets, etc.
In addition, here we discuss some practical issues that need to be considered when conducting fuzzing experiments.
First, it is important to monitor the operating status of a fuzzing experiment such as CPU utilization to determine whether the experiment is executed normally. 
In general, 
if the CPU utilization rate is low (e.g., less than 80\%), 
it may indicate
that the fuzzing status is abnormal. 
For instance, 
when there is a large amount of disk I/O operations, 
CPU has to wait for these operations before it does real fuzzing work.
Second, 
it is important to mitigate unnecessary disk I/O operations, 
especially when conducting many fuzzing experiments
on a sever simultaneously, 
where disk I/O can easily become the bottleneck. 
For instance,
the target program may output a large amount of new files during the fuzzing process, which may cause heavy disk output operations, while these output files are not important for evaluations.
In this situation, it is suggested using a RAM disk or not saving the output files generated by the target benchmark program.

\subsection{The Benchmarks for Evaluating Fuzzers}

The current fuzzing benchmarks are not satisfactory \cite{klees2018evaluating}. Considering the practical usability issue, we construct a pragmatic benchmark suite which consists of 20 real-world programs at the current version and has the following advantages.
First, the \system benchmark programs have various expressiveness in functionality, size, vulnerability, etc., which can provide comprehensive evaluations on fuzzers and can better reflect a fuzzer's performance on the real-world programs.
Second, the \system  benchmark can be used to provide more objective and fairer evaluations.
As shown in Section \ref{general}, no fuzzer outperforms the others on all programs, which somehow demonstrates the bias and subjectivity in many existing fuzzing papers.
Third, different from traditional benchmark design methods that 
inject artificial bugs, our method does not change the original code of the real-world programs in order to keep its raw features, but focusing on providing convenient offline result analysis methods.
Specifically, for each program, we provide crash analysis methods including crash triage, CVE matching, bug severity analysis, etc.
Therefore, the \system benchmark is easily usable like the synthetic benchmarks. 
In addition, to the best of our knowledge, we are the first to construct the \emph{CVE keywords database} which greatly reduces the human labors in CVE matching.

Note that fuzzing benchmarks need to be updated and improved with the development of fuzzers, and there still needs more research on designing benchmarks.
That is why we design \system as an open-source and extensible platform. 
There are still limitations with the \system benchmark and can be improved and extended from many perspectives.
First, in this paper, we select the programs mainly from the top fuzzing papers.
In the future, there are many other resources such as  vulnerability-related websites \cite{exploit-db, CVE-details, CVE, Securityfocus, Securitytracker} that can be leveraged to select programs.
Second, the current \system benchmark  mainly focuses on the general program-level fuzzers.
It would be better to incorporate \system with more benchmarks for other types of fuzzers such as compiler fuzzers  and kernel fuzzers.

\subsection{Performance Metrics}
The existing metrics are rough and  incomprehensive.
To solve the problem,
\system provides six categories of metrics which aim to provide more comprehensive evaluations on fuzzers.
Here we discuss the limitation of these metrics along with the related  interesting research questions which can be considered as the future work.
First is the categories of the metrics.
In this paper we classify the metrics into six categories.
It calls for more research to provide a more reasonable classification.
Second, it needs more research on the concrete metrics of each category.
For instance, we use the CVSS score and the \texttt{Exploitable} tool to evaluate the severity of the bugs.	
However, each individual metric has its own limitation.
As the CVSS score takes multiple factors  (e.g., attack complexity and required privileges) into consideration,  the single numeric score may not be able to accurately reflect the impact of each individual perspective.
\texttt{Exploitable} determines the severity of a bug based on a list of rules, whose accuracy may be affected by the rationality of the rules.
Thus, the choice of concrete metrics to evaluate the severity of bugs should be updated when better standards/methods are proposed.
In addition, it needs more theoretical research on the metrics.
For instance, when conducting statistical test on the number of unique bugs, 
we can only use non-parametric statistical methods such as the Mann-Whitney U test, which makes no assumption on the distribution of the population.
It is interesting to study the distribution of the number of bugs in multiple repetitions and provide more suitable metrics to assess it.
Third,
it is necessary to study the priority of each metric.
In our opinion, 
the number of bugs and the quality of bugs are more important than the stability of finding bugs,
as finding less or trivial bugs stably is much less valuable than finding more high-risk bugs occasionally.
Fourth, 
it could be desirable to design a scoring method which combines different metrics to generate a conclusive numerical score for assessing a fuzzer's  performance.

\section{Conclusion}\label{conclusion}
In this paper, 
we propose and implement \system,
an open-source, holistic, and pragmatic metrics-driven platform for evaluating fuzzers in a comprehensive and fair manner.
\system has incorporated 35 fuzzers,
20 real-world benchmark programs,
and six categories of performance metrics.
We test the usability of the 35 fuzzers and discover a number of flaws.
Leveraging \system,
we systematically compare the state-of-the-art fuzzers.
Based on the experimental results,
we have the following important observations.
First,
no fuzzer always performs better than others,
revealing potential subjectivity and bias in the evaluations of existing fuzzing works.
Second,
the performance of fuzzers on the synthetic benchmark programs may not be consistent with that on the real-world programs,
which confirms the importance of using pragmatic benchmark programs.
Third,
the performance of fuzzers varies with different performance metrics,
which indicates that the fuzzers need to be evaluated with more comprehensive performance metrics for reliable assessments.
In addition,
we identify new factors such as \emph{instrumentation methods} and \emph{crash analysis tools}  that can significantly affect the evaluation of fuzzers.
We have made \system publicly available to facilitate future fuzzing research.

\section*{Acknowledgments}
We sincerely appreciate the anonymous reviewers for their valuable comments to improve our paper.
This work was partly supported by the National Key Research and Development Program of China under No. 2018YFB0804102 and No. 2020YFB1804705,
NSFC under No. U1936215, U1836202, and 61772466,
the Zhejiang Provincial Natural Science Foundation for Distinguished Young Scholars under No. LR19F020003,
the Zhejiang Provincial Key R\&D Program under No. 2019C01055,
the Fundamental Research Funds for the Central Universities (Zhejiang University NGICS Platform),
the Ant Financial Research Funding,
the Industrial Internet innovation and development project under No. TC190A449,
the Key Research and Development Program of Zhejiang Province under No. 2020C01021,
and Major Scientific Project of Zhejiang Lab under No. 2018FD0ZX01.
Peng Cheng's research was partly supported by NSFC under grant 61833015.
Ting Wang's research was partly supported by the National Science Foundation under Grant No. 1953893, 1953813, and 1951729.
Wei-Han Lee's research was sponsored by the U.S. Army Research Laboratory and the U.K. Ministry of Defence under Agreement Number W911NF-16-3-0001. The views and conclusions contained in this document are those of the authors and should not be interpreted as representing the official policies, either expressed or implied, of the U.S. Army Research Laboratory, the U.S. Government, the U.K. Ministry of Defence or the U.K. Government. The U.S. and U.K. Governments are authorized to reproduce and distribute reprints for Government purposes notwithstanding any copyright notation hereon.

\bibliographystyle{IEEEtran}
\bibliography{myref}

\end{document}